\documentclass[superscriptaddress, amsmath,amssymb, pre,twocolumn]{revtex4-1}

\usepackage{graphicx}
\usepackage{dcolumn}
\usepackage{bm}
\usepackage{hyperref}
\usepackage{color}
\usepackage{setspace}

\begin{document}

\title{Phase behavior and morphology of multicomponent liquid mixtures}

\author{Sheng Mao}
\affiliation{Department of Mechanical and Aerospace Engineering, Princeton University, Princeton, NJ, 08544}
\author{Derek Kuldinow}
\affiliation{Department of Mechanical and Aerospace Engineering, Princeton University, Princeton, NJ, 08544}
\affiliation{Department of Mechanical Engineering and Materials Science, Yale University, New Haven, CT 06511}
\author{Mikko P. Haataja}
\affiliation{Department of Mechanical and Aerospace Engineering, Princeton University, Princeton, NJ, 08544}
\affiliation{Princeton Institute for the Science and Technology of Materials (PRISM), Princeton University, Princeton, NJ 08544, USA}%
\author{Andrej Ko\v smrlj}%
\affiliation{Department of Mechanical and Aerospace Engineering, Princeton University, Princeton, NJ, 08544}
\affiliation{Princeton Institute for the Science and Technology of Materials (PRISM), Princeton University, Princeton, NJ 08544, USA}%

\date{\today}

\begin{abstract}
Multicomponent systems are ubiquitous in nature and industry. While the physics of few-component liquid mixtures (i.e., binary and ternary ones) is well-understood and routinely taught in undergraduate courses, the thermodynamic and kinetic properties of $N$-component mixtures with $N>3$ have remained relatively unexplored. An example of such a mixture is provided by the intracellular fluid, in which protein-rich droplets phase separate into distinct membraneless organelles. In this work, we investigate equilibrium phase behavior and morphology of $N$-component liquid mixtures within the Flory-Huggins theory of regular solutions. In order to determine the number of coexisting phases and their compositions, we developed a new algorithm for constructing complete phase diagrams, based on numerical convexification of the discretized free energy landscape. Together with a Cahn-Hilliard approach for kinetics, we employ this method to study mixtures with $N=4$ and $5$ components. We report on both the coarsening behavior of such systems, as well as the resulting morphologies in three spatial dimensions.  We discuss how the number of coexisting phases and their compositions can be extracted with Principal Component Analysis (PCA) and K-Means clustering algorithms. Finally, we discuss how one can reverse engineer the interaction parameters and volume fractions of components in order to achieve a range of desired packing structures, such as nested ``Russian dolls'' and encapsulated Janus droplets.
\end{abstract}

\maketitle

\section{Introduction}

Phase separation and multi-phase coexistence are commonly seen in our everyday experience, from simple observations of the demixing of water and oil to sophisticated liquid extraction techniques employed in the chemical engineering industry to separate the components of solutions. In non-biological systems, phase separation has been studied for a long time dating back to Gibbs~\cite{gibbs1906}.  Very recently, it has been demonstrated that living cells are also multicomponent mixtures composed of a large number of components, with phase separation behavior reminiscent of those found in inanimate systems in equilibrium~\cite{runnstrom1963sperm,walter1995phase,muschol1997liquid,iborra2007can,weber2015inverse,molliex2015phase,Shin2017}. 
This process has been shown to drive the formation of membraneless organelles in the form of simple droplets~\cite{brangwynne2009germline,brangwynne2011active,berry2015rna,molliex2015phase,patel2015liquid,mitrea2016nucleophosmin, Shin2017}, and even hierarchical nested packing structures~\cite{feric2016}.  

The physics of binary ($N=2$) and ternary ($N=3$) mixtures are well-understood by now, with binary mixtures comprising standard material in undergraduate statistical thermodynamics courses. Given, say, a molar Gibbs free energy of the mixture as a function of composition, the presence of coexisting phases can be ascertained via the common tangent construction, and repeating this process at several temperatures, the phase diagram can be readily constructed. Similar arguments also hold for ternary \cite{jones2002soft,porter2009phase} and $N>3$ mixtures, while the construction of phase diagrams becomes rapidly more challenging, in accordance with the Gibbs phase rule~\cite{gibbs1906}, which states that the maximum number of coexisting phases in an $N$-component mixture is $N+2$. On the other hand, when $N \gg 1$, statistical approaches for predicting generic properties of phase diagrams become applicable. 

In their pioneering work, Sear and Cuesta \cite{sear2003} modeled an $N$-component system with $N \gg 1$ within a simple theoretical approach, which incorporated entropy of mixing terms and interactions between the components at the level of second virial coefficients; the virial coefficients were in turn treated as Gaussian random variables with mean $b$ and variance $\sigma^2$. In the special case of an equimolar mixture, their analysis based on Random Matrix Theory showed that for $N^{1/2} b/\sigma < -1$, the system is likely to undergo phase separation via spinodal decomposition, leading to compositionally distinct phases. On the other hand, for $N^{1/2} b/\sigma > -1$, the mixture will likely undergo a condensation transition, which leads to the formation of two phases differing in only density (and not relative compositions). These predictions were later confirmed by Jacobs and Frenkel using grand canonical Monte Carlo simulations of a lattice gas model with up to $N=16$ components \cite{jacobs2013}. In subsequent work \cite{jacobs2017}, Jacobs and Frenkel argued that multiphase coexistence in biologically-relevant mixtures with $N \gg 1$ does not result from the presence of a large number of components, but requires fine tuning of the intermolecular interactions.

In order to begin to bridge the gap between the well-studied binary and ternary systems on the one hand, and mixtures with $N \gg 1$ on the other,  herein we systematically investigate the phase behavior and morphology of liquid mixtures with $N=4$ and $5$ components. We develop a method to construct full phase diagrams of such systems based on free energy convexification within the Flory-Huggins theory of regular solutions \cite{flory1942,huggins1941}, and employ the Cahn-Hilliard \cite{cahn1958} formalism to study associated domain growth and coarsening processes during morphology evolution. In order to identify and locate the emerging phases in the simulations, we employ a combination of principal component analysis (PCA)\cite{jolliffe2011principal} and K-Means clustering method \cite{hartigan1979algorithm} to translate local compositions to phase indicator functions. The phase indicator functions are, in turn, employed to quantify the domain growth and coarsening kinetics. Specifically, characteristic domain sizes for each phase are extracted from time-dependent structure factors, and their behavior is compared against classical theories of coarsening kinetics \cite{lifshitz1961kinetics, bray1989exact, bray2002theory,berry2018physical}. Finally, we demonstrate how tuning the interfacial energies between phases enables one to engineer morphologies with a wide range of packing structures, including Janus-particle like domains and nested ``Russian doll'' droplets-within-droplets with up to 5 layers.

The rest of this paper is organized as follows. In Section \ref{sec:phase_diagram}, the equilibrium phase behavior of an $N$-component liquid mixture is examined within the Flory-Huggins (F-H) theory of regular solutions. An algorithm based on convex hull construction to compute the phase diagram of the mixture is developed, and a graph theory based method is employed to determine the number of coexisting phases corresponding to given set of interaction parameters within the F-H theory and average composition. In Section \ref{sec:kinetics}, the Cahn-Hilliard formalism is employed to both incorporate interfacial effects within the F-H theory and model the spatio-temporal evolution kinetics of the local compositions. The method to construct local phase indicator functions is also outlined in this section.  Resulting microstructures for representative $4$-component systems are presented in turn in Section \ref{sec:4}. We also demonstrate how interaction parameters can be tuned to achieve different final packing morphologies of the coexisting phases.  In Section \ref{sec:5}, 
we focus on the coarsening kinetics of the phase separation process. We examine the validity of the dynamic scaling theory in multicomponent settings and discuss the coarsening behavior due to the multiple coexisting phases. In Section \ref{sec:design}, we provide guidelines for the design of hierarchical nested structures, and employ them to design three different nested structures in $5$-component mixtures. Finally, brief concluding remarks can be found in Section \ref{sec:conc}.  
 
\section{Equilibrium phase behavior}
\label{sec:phase_diagram}

\subsection{Flory-Huggins theory}
In this study, we model a dilute solution as a continuum multicomponent incompressible fluid composed of $N$ different components, where $\phi_{i}$ represents the volume fraction of component $i$ ($\sum_{i=1}^N \phi_i=1$). For simplicity, we only focus on the phase behavior of condensates and solvent is not explicitly considered in our treatment.

\begin{figure*}[!t]
\centering
  \includegraphics[height=9cm]{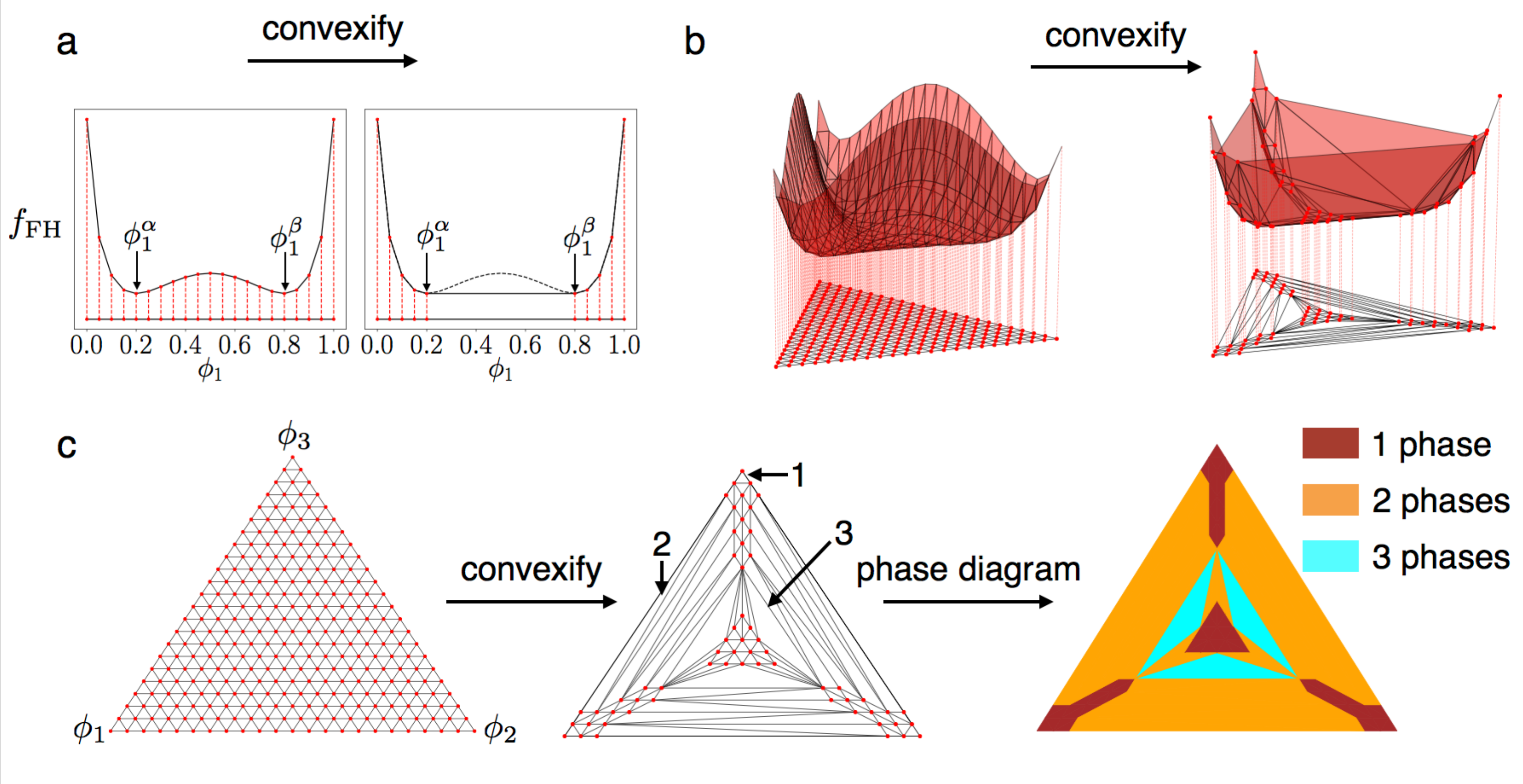}
  \caption{Construction of phase diagrams based on finding the convex hulls of free energy functions for (a) binary and (b-c) ternary mixtures. (a) The original free energy function (black solid line in left) and the convexified one using the common tangent construction (black solid line in right) for a binary mixture with Flory interaction parameter $\chi_{12} = 2.31$. Red dots and lines correspond to a discrete approximation of the free energy function evaluated on a uniform grid (left) and to the convex hull of the free energy function (right). Red points are projected to the abscissa (composition space). Short projected segments from the convex hull correspond to single phase regions, while long projected segments correspond to two phase ones. (b) Discrete approximation of the free energy function (left) and its convex hull (right). (c) Projected triangles from the original free energy function (left) and from the convex hull (middle). The number of stretched sides for projected triangles corresponds to the number of coexisting phases for the composition points that reside within such triangles. This information is used to construct the ternary phase diagram (right). }
  \label{fig:convex-construct}
\end{figure*}

First, we briefly review some properties of binary mixtures. According to the Flory-Huggins theory of regular solutions \cite{flory1942,huggins1941} the free energy density (per volume) is expressed as
\begin{equation}
\label{eqn:binary-FH}
f_\textrm{FH} (\phi_1, \phi_2) = {c_0RT}\left[ {\phi_1} \ln \phi_1 + {\phi_2} \ln \phi_2 + \phi_1 \phi_2 \chi_{12}\right],    
\end{equation}
where $c_0$ is the total molar concentration of solutes, $R$ is the gas constant, $T$ the absolute temperature, and $\phi_1+\phi_2=1$ due to incompressibility.  The first two terms in Eqn.~(\ref{eqn:binary-FH}) incorporate the entropy of mixing, which favors a homogeneous binary mixture. The last term describes the enthalpic part of the free energy. The Flory interaction parameter $\chi_{12}$ is related to the pairwise interaction energies $\omega_{ij}$ between components $i$ and $j$ as $\chi_{12} = z(2\omega_{12} - \omega_{11} - \omega_{22})/(2k_BT)$, where $k_B$ denotes the Boltzmann constant and $z$ is the coordination number\cite{cahn1958}. When $\chi_{12} < 0$, the two different components attract each other and favor mixing. When $\chi_{12} > 0$, the two components repel each other, which can drive the system to demix and form two coexisting phases (one enriched with component 1 and one enriched with component 2) once the Flory parameter becomes sufficiently large  ($\chi_{12}>2$), such that enthalpy dominates over the mixing entropy \cite{porter2009phase,jones2002soft} (see Fig.~\ref{fig:convex-construct}a).

The Flory-Huggins free energy density in Eq.~(\ref{eqn:binary-FH}) can be easily generalized to describe an incompressible liquid mixture with $N$ different components as \cite{berry2018physical}
\begin{equation}
\label{eqn:multi-FH}
f_{FH} (\{\phi_i\}) =c_0RT  \left[ \sum_{i=1}^{N} {\phi_i} \ln \phi_i  + \frac{1}{2}\sum_{i,j =1}^{N} \phi_i \phi_j \chi_{ij}\right]. 
\end{equation}
The first term describes the mixing entropy and the second term describes the enthalpic part, where  $\chi_{ij}=z(2\omega_{ij} - \omega_{ii} - \omega_{jj})/(2k_BT)$ are the Flory interaction parameters between components $i$ and $j$. Note that by definition $\chi_{ii}=0$. 

The Flory-Huggins theory presented above has been widely used to model mixtures of regular solutions in dilute limit and was also generalized to model polymeric systems~\cite{sariban1987critical,favre1996application,glotzer1995computer,lapena1999effect,berry2018physical}. Now, according to the Gibbs phase rule~\cite{gibbs1906}, there can be as many as $N$ coexisting liquid phases at fixed temperature and pressure, but the actual number depends on the interaction parameters $\{\chi_{ij}\}$ and the average composition $\{\overline \phi_i\}$.
In the next subsection we describe an algorithm for constructing a complete phase diagram for a given set of interactions $\{\chi_{ij}\}$, which is based on the convexification of the free energy density in Eq.~(\ref{eqn:multi-FH}).

\subsection{Phase diagram based on the convex hull construction}

In order to construct a phase diagram, one needs to find the convex envelope of the free energy density in Eq.~(\ref{eqn:multi-FH}). For binary mixtures the free energy density depends on a single variable ($\phi_1$) and the two phase coexistence regions can be identified via the standard common tangent construction~\cite{porter2009phase,jones2002soft}. For mixtures with $N$ components, the free energy landscape can be represented as an $(N-1)$-dimensional manifold embedded in an $N$-dimensional space. The regions in composition space that correspond to the $P$ coexisting phases can in principle be obtained by identifying common tangent hyperplanes that touch the free energy manifold at $P$ distinct points. This is a very daunting task for mixtures with many components. Here we describe how a complete phase diagram can be obtained via a convex hull construction of the discretized free energy manifold. This method was initially introduced by Wolff et al.~\cite{wolff2011} for the analysis of ternary mixtures, and here we generalize it to systems with an arbitrary number of components $N$. 

To illustrate the main idea of the algorithm, it is useful to first recall the phase diagram construction process for binary mixtures ($N=2$). When the Flory parameter is sufficiently large ($\chi_{12}>2$), the free energy density becomes a double well potential with two minima located at $\phi_1^\alpha$ and $\phi_1^\beta$ (see Fig.~\ref{fig:convex-construct}a). When the average composition $\overline \phi_1$ is between the two minima ($\phi_1^\alpha < \overline \phi_1 < \phi_1^\beta$), the free energy of the system can be lowered by demixing and thus forming two phases $\alpha$ and $\beta$ with compositions $\phi_1^\alpha$ and $\phi_1^\beta$, respectively. The volume fractions $\eta_\alpha$ and $\eta_\beta$ ($\eta_\alpha+\eta_\beta=1$) of the two phases can then be obtained from the lever rule~\cite{porter2009phase,jones2002soft}, such that $\overline{\phi}_1 = \eta_{\alpha} \phi_1^{\alpha} + \eta_{\beta} \phi_1^{\beta}$.

\begin{figure*}[!t]
\centering
  \includegraphics[width=0.95\textwidth]{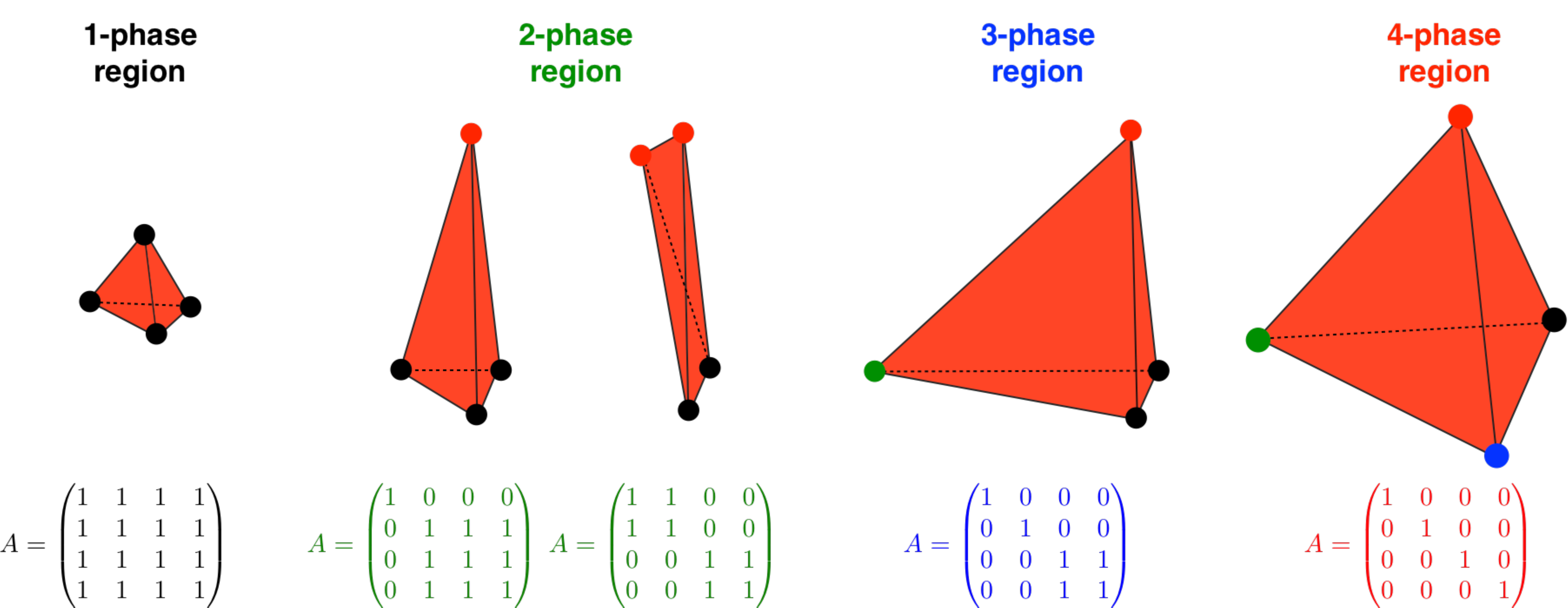}
  \caption{Distinct types of stretched tetrahedra, which correspond to regions with different numbers of coexisting phases, resulting from the projection of the free energy convex hull to the composition space, and their respective adjacency matrices $A$ (see text). Vertices with identical colors which are connected with short line segments correspond to the same phase, while vertices with opposite colors that are connected with long line segments correspond to different phases.}
  \label{fig:stretched-tetrehedron}
\end{figure*}

Now, we show how identical information can be obtained via the convex hull construction of the discretized free energy landscape. First, we discretize the composition space $\phi_1 \in [0,1]$ with regular segments and make a discrete approximation of the free energy function~(see Fig.~\ref{fig:convex-construct}a). Then, we construct the convex hull of the discretized free energy function and we project it back onto the composition space $\phi_1 \in [0,1]$. Note that the projected segments remain unchanged in the regions that correspond to a single phase (i.e. for $\overline \phi_1 < \phi_1^\alpha$ and $\overline \phi_1 > \phi_1^\beta$), while all the segments between the two free energy minima are replaced with a single {\it{stretched}} line segment. The stretched segment of the projected free energy convex hull thus denotes the two phase coexisting region, where the two ends of the segment correspond to the compositions $\phi_1^\alpha$ and $\phi_1^\beta$ of the two coexisting phases, respctively. Note that the discretized points may not exactly coincide with the true free energy minima, but the error can be made arbitrarily small by refining the mesh.

For a ternary mixture we follow the same procedure. First, we discretize the composition space with small equilateral triangles and we make a discrete approximation of the free energy function~(see Fig.~\ref{fig:convex-construct}b). Then, we construct the convex hull of the discretized free energy function and project it back onto the composition space. Now, there are in general three different types of projected triangles: triangles with three short sides, triangles with two elongated sides, and triangles with three elongated sides (see Fig.~\ref{fig:convex-construct}c). According to Wolff et al.\cite{wolff2011}, the three different types of triangles correspond to a single phase regions, 2-phase regions, and 3-phase regions, respectively. For the 3-phase region, the corners of triangles describe the equilibrium compositions $\{\phi_i^{\alpha}\}$ of the phases, where $i$ and $\alpha$ denote the indices of the component and of the phase, respectively. For a mixture with average composition $\{\overline \phi_i\}$, that resides inside such a triangle, the mixture phase separates into three phases with volume fractions $0<\eta_\alpha<1$, such that $\overline \phi_i = \sum_\alpha \eta_{\alpha} \phi_i^{\alpha}$ and $\sum_\alpha \eta_\alpha = 1$. For the 2-phase regions, the two long sides of a projected triangle are approximations for the tie-lines that connect the two coexisting phases, while the two corners that are connected by the short side correspond to the identical phase. For a mixture with an average composition that lies inside such a triangle, the mixture phase separates into the two phases located at the ends of the tie-line. Refining this process with arbitrarily small mesh sizes enables one to obtain complete information about ternary phase diagrams\cite{wolff2011}~(see Fig.~\ref{fig:convex-construct}c).

Conceptually, it is straightforward to generalize the phase diagram construction to mixtures with $N>3$ components. First, we discretize the composition space with regular $(N-1)$-dimensional simplexes and make a discrete approximation of the free energy function. Then, we construct the convex hull of the discretized free energy function and project it back onto the composition space. The projected $(N-1)$-dimensional simplexes are distorted when they correspond to regions with multiple coexisting phases. In Fig.~\ref{fig:stretched-tetrehedron} we display all distinct types of distorted tetrahedra (simplexes) for an $N=4$ component mixture. Next we demonstrate that determining the number $P$ of different coexisting phases for distorted simplexes can be mapped to the problem of counting the number of distinct connected components in a simple graph.

To this end, based on our knowledge from ternary mixtures~(see Fig.~\ref{fig:convex-construct}c), we make the observation that the two vertices of simplexes that are connected by a \emph{stretched} line segment correspond to two distinct phases, while the two vertices that are connected by a \emph{short} line segment correspond to the identical phase. Now, the vertices of simplexes can be represented as graph vertices. The two simplex vertices $i$ and $j$ are considered connected (disconnected), i.e.~they correspond to the identical phase (two distinct phases), when their Euclidian distance $||\vec{r}_i - \vec{r}_j||$ in the composition space is smaller (larger) than some threshold $\Delta$, which we typically set to be slightly larger than the initial mesh size. Note that the threshold needs to be slightly larger, because the convex hull algorithm may return small irregular simplexes in the $1$-phase regions (see Fig.~\ref{fig:convex-construct}c). In practice we find that the threshold $\Delta$ needs to be set at about $\sim5$ times the initial mesh size. Thus we define the adjacency matrix $A_{ij}$ for such graph as
\begin{equation}
A_{ij} = 
\begin{cases}
1, \quad ||\vec{r}_i - \vec{r}_i|| \leq \Delta,  \\
0, \quad \text{otherwise.}
\end{cases}
\end{equation}

\begin{figure}[!t]
\centering
  \includegraphics[width=0.5\textwidth]{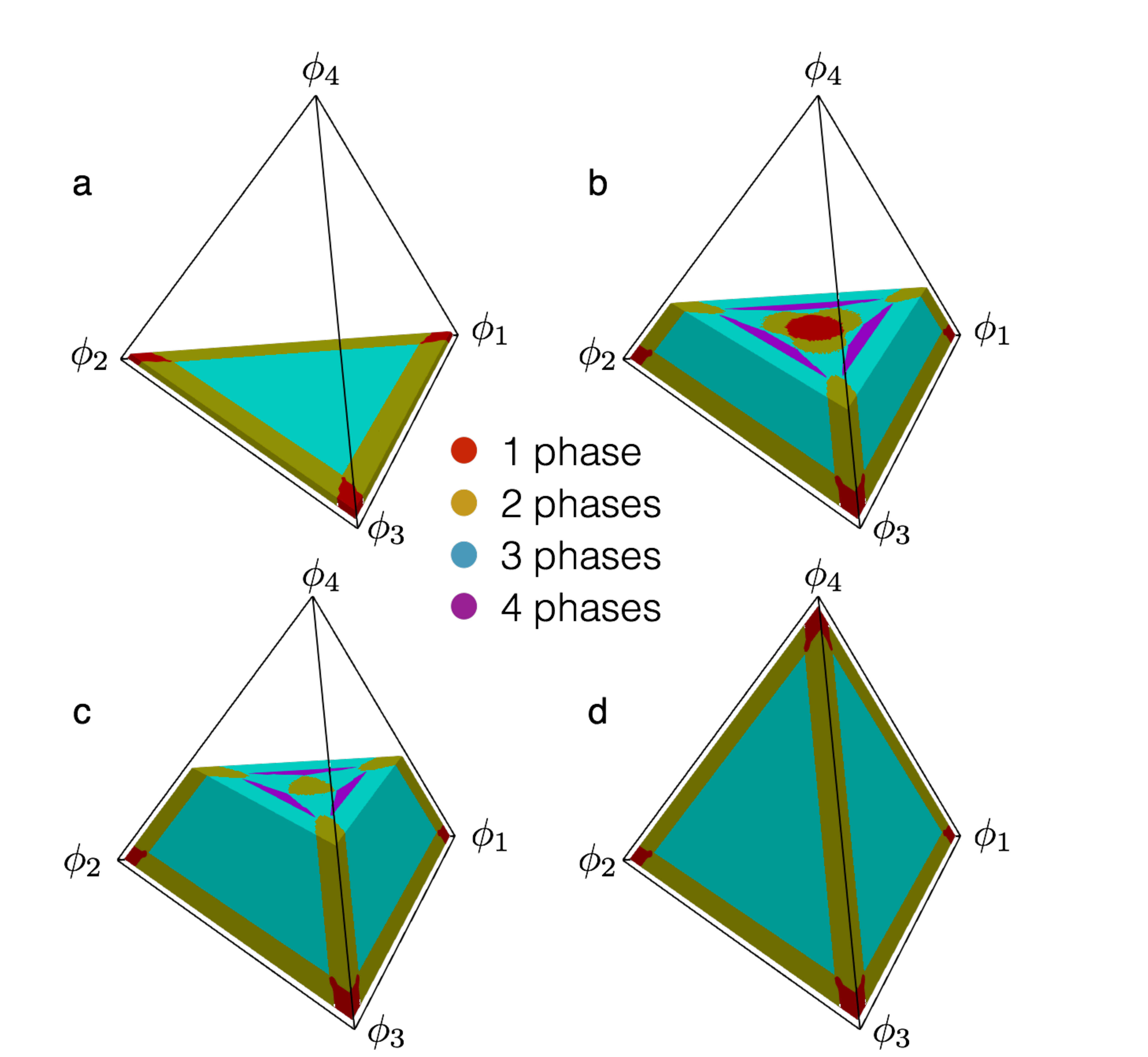}
  \caption{Phase diagram for a 4-component mixture with symmetric interactions $\chi_{ij}\equiv 3.10$ for $i\ne j$. From (a) to (d) the slicing planes are at $\phi_4 = 0.025$, $0.25$, $0.50$, and $1.0$, respectively.}
  \label{fig:4-component-phase}
\end{figure}

The number of distinct phases $P$ for a given simplex is thus equivalent to determining the number of distinct connected components for a graph characterized with the adjacency matrix $A_{ij}$. From graph theory~\cite{chung1997spectral} we know that this number is related to the spectrum of the Laplacian of $A$, which is defined as $L_{ij} = D_{ij} - A_{ij}$, where $D_{ij}$ is the weight matrix defined as $D_{ii} = \sum_{k} A_{ik}$ and $D_{ij}=0$ for $i \ne j$. The number of distinct connected components is then equal to the algebraic multiplicity of the eigenvalue 0. 
The examples for tetrahedra in $N=4$ component mixtures are shown in Fig.~\ref{fig:stretched-tetrehedron}.
The locations of vertices also provide approximate values for the compositions $\{\phi_i^\alpha\}$ of each phase $\alpha$.  For a mixture with an average composition $\{\overline \phi_i\}$ that resides inside a simplex, the mixture phase separates into $P$ coexisting phases with volume fractions $0<\eta_\alpha<1$, such that $\overline \phi_i = \sum_{\alpha=1}^P \eta_{\alpha} \phi_i^{\alpha}$. The volume fractions $\lbrace \eta_\alpha \rbrace$ can be determined by calculating the pseudo-inverse~\cite{moors1920reciprocal,penrose1955generalized} of the $N\times P$ matrix ${\Phi}\equiv \phi_i^\alpha$ as
\begin{equation}
\label{eq:volume_fractions}
\eta_{\alpha} = R_{\alpha j} \overline\phi_{j}, \ \ {\bf R} = \left( { \Phi}^T { \Phi}  \right)^{-1} {\Phi}^T.
\end{equation}
With this procedure we were able to construct phase diagrams for mixtures with $N=4$ (see Fig.~\ref{fig:4-component-phase} and \href{http://www.princeton.edu/~akosmrlj/videos/}{Video~S1}) and $N=5$ components, where we used the standard Qhull algorithm to construct convex hulls.~\cite{barber1996quickhull}

The algorithm described above is general and can in principle be used to construct phase diagrams for mixtures with an arbitrary number $N$ of components with a given set of interaction parameters $\lbrace \chi_{ij} \rbrace$. However, it is practically impossible to use this procedure for constructing phase diagrams for mixtures with $N>5$ components, which can be demonstrated by considering the scaling of computational complexity. First, we need to generate a uniform mesh for an $N-1$ dimensional simplex to discretize the composition space~(see Fig.~\ref{fig:convex-construct}c). The number of different points scales as $M_p \sim M^{(N-1)}$, where $M$ is the number of grid points for each component. In order to precisely capture the boundaries between different regions on a phase diagram, one has to use sufficiently fine mesh ($M \gg 1$) of the discretized composition space. For $N>3$ the computational time of the Qhull algorithm scales as $\mathcal{O} (M_p f_v/M_v)$, where $M_v\le M_p$ is the number of vertices on the convex hull and $f_v$ is the maximum number of facets for a convex hull of $M_v$ vertices \cite{barber1996quickhull}. We note that the number of facets grows as $f_v \sim M_v^{\lfloor N/2 \rfloor}/\lfloor N/2 \rfloor!$, where $\lfloor \cdot \rfloor$ is the floor function. This means that, in the worst case scenario, the computational complexity scales as $\mathcal{O} (M^{(N-1)\lfloor N/2 \rfloor})$, when the free energy landscape is convex to begin with ($M_v=M_p$). In practice, we managed to use this algorithm to construct phase diagrams for mixtures with up to $N=5$ components.

\section{Phase separation kinetics: Cahn-Hilliard formalism and microstructural characterization}
\label{sec:kinetics}

\subsection{Cahn-Hilliard formalism}
The convex hull algorithm described in the previous section can predict the number of coexisting phases, but it cannot provide any information about the equilibrium microstructure, which is governed by the interfacial properties between phases. To account for such effects, we follow the treatment of Cahn-Hilliard~\cite{cahn1958}.

With regard to kinetics, Hohenberg and Halperin~\cite{hohenberg1977theory} introduced several standard dynamic models of domain growth and phase separation processes. The form of the dynamic evolution equations depends on the nature of the order parameter (conserved or non-conserved) and the physics of the problem (e.g., relative importance of diffusive vs.~advective transport processes). Such models have been successfully employed to study a wide spectrum of problems in materials science, e.g.~solidification, spinodal decomposition and many others~\cite{ratke2002growth}.
Recently, these models have also been used to study compositional domain formation in lipid bilayer membranes \cite{camley2010dynamic,han2013compositional,stanich2013coarsening}. Several different models have been proposed for the investigation of multicomponent multiphase systems~\cite{hoyt1990continuum,cogswell2011thermodynamic}. In this paper we follow the treatment by Cahn and Hilliard~\cite{cahn1958} to investigate phase separation of $N=4$ and $N=5$ component mixtures in three spatial dimensions.

Before writing the general expression incorporating interfacial effects for an $N$--component mixture, it is useful to briefly comment on binary mixtures. For such systems, Cahn and Hilliard postulated the free energy density
$f(\phi_1, \nabla \phi_1) = f_{\text{FH}}(\phi_1,1-\phi_1) + c_0RT \chi_{12}\lambda_{12}^2 (\nabla \phi_1)^2$, where the first term describes the Flory-Huggins part of the free energy in Eq.~(\ref{eqn:binary-FH}) and the second term is related to the interfacial energy of the system. Here, $\lambda_{12}$ is related to the characteristic width of the interface (usually taken proportional to the range of interaction between molecules~\cite{cahn1958}). Note that a stable interface can form only when $\chi_{12}>0$. The Cahn-Hilliard approach assumes that the interfacial energies are primarily coming from enthalpic interactions. However, for long-chain polymers the entropic effects may become important, and for such systems the interfacial part can be described within the Flory-Huggins-de Gennes approach\cite{deGennes1980dynamics}. 

In order to generalize the Cahn-Hilliard formalism to mixtures with $N$ components, it is useful to first rewrite the Cahn-Hilliard free energy density for a binary mixture in a symmetric form as
\begin{eqnarray}
f &=& {c_0RT}\left[ {\phi_1} \ln \phi_1 + {\phi_2} \ln \phi_2 + \chi_{12} \phi_1 \phi_2  \right.\nonumber \\
&& \quad \quad \quad \left.- \lambda_{12}^2 \chi_{12}\nabla \phi_1 \cdot \nabla \phi_2 \right],
\end{eqnarray}
where $\phi_1 + \phi_2 = 1$, and thus $\nabla \phi_1 + \nabla \phi_2 = 0$. The negative sign before the $\lambda_{12}^2$ is thus merely a consequence of incompressibility, while interfacial stability still requires that $\chi_{12} > 0$. Following the procedure documented in Ref.~\cite{cahn1958}, we can generalize the above free energy density to an $N$-component mixture as
\begin{eqnarray}
f  &=& c_0RT \left[ \sum_{i=1}^{N} {\phi_i} \ln \phi_i  + \frac{1}{2}\sum_{i, j=1}^{N} \chi_{ij} \phi_i \phi_j  \right. \nonumber \\
&& \quad \quad \quad \left.    - \frac{\lambda^2}{2} \sum_{i, j=1}^{N} \chi_{ij} \nabla \phi_i \nabla \phi_j
\right],
\label{eqn:CH-multi}
\end{eqnarray}
where, for simplicity, we assumed that the ranges of intermolecular interactions are identical such that $\lambda_{ij} \equiv \lambda$. The parameter $\lambda$ thus describes the characteristic width of all interfaces in the system.

Now, the equilibrium packing (i.e., morphology) of coexisting phases can in principle be obtained by minimizing the total free energy functional
\begin{equation}
F =  \int_V \! d^3\vec{r} \ f\left[\lbrace \phi_i(\vec{r}), \nabla \phi_i(\vec{r}) \rbrace\right],
\label{eq:tot_free_energy}
\end{equation}
subject to the fixed average composition $\overline \phi_i = \int_V d^3\vec{r}\ \phi_i(\vec{r})$. 
This is in general a very hard optimization problem, but one can learn much about the local microstructure by considering the interfacial energies (also called surface tensions) $\gamma_{\alpha \beta}$ between different phases. According to Cahn and Hilliard \cite{cahn1958}, the interfacial energy between the two phases $\alpha$ and $\beta$ with compositions $\lbrace \phi_i^\alpha\rbrace $ and $\lbrace \phi_i^\beta\rbrace $, respectively, can be estimated as
\begin{equation}
\label{eqn:sigma-cal}
\gamma_{\alpha \beta} \approx 2\lambda c_0RT 
\int_{0}^{1} \! d\eta \, \sqrt{\kappa_{\alpha\beta} \, \Delta \tilde{f}_{FH}(\eta)} ,
\end{equation}
where $\eta$ is a parameter that interpolates between the two phases such that $\phi_i = \eta \phi_i^\alpha + (1-\eta) \phi_i^\beta$.  The other two quantities in Eqn.(\ref{eqn:sigma-cal}) are defined as  $\kappa_{\alpha \beta} = - \frac{1}{2}\sum_{i,j} \chi_{ij}(\phi_i^\alpha - \phi_i^\beta) (\phi_j^\alpha - \phi_j^\beta)$, and $\Delta \tilde{f}_{FH}(\eta) = \tilde{f}_{FH}(\phi_i) - \eta \tilde{f}_{FH}(\phi_i^{\alpha}) - (1-\eta)\tilde{f}_{FH}(\phi_i^{\beta})$, where $\tilde{f}_{FH} = f_{FH}/(c_0RT)$. Note that the interface between phases $\alpha$ and $\beta$ is stable only when $\kappa_{\alpha\beta}>0$, due to the fact that the excess free energy $\tilde{f}_{FH}(\eta)$ is always positive. This observation provides some restrictions for the Flory interaction parameters $\lbrace \chi_{ij} \rbrace$, that must satisfy relation
$\sum_{i,j=1}^{N} a_i \chi_{ij} a_j <0$ for any $\lbrace a_i\rbrace$ with $\sum_i a_i=0$. 

Here we briefly comment on the special case, where each of the two phases $\alpha$ and $\beta$ are predominantly composed of components $a$ and $b$, respectively, i.e.~$\phi_i^\alpha \approx \delta_{ia}$ and $\phi_i^\beta \approx \delta_{ib}$, where $\delta_{ij}$ denotes the Kronecker delta. This special case typically occurs when Flory interaction parameters are large ($\chi_{ij} \gg 1$). For this special case, it is easy to show that $\Delta \tilde{f}_{FH} \approx \chi_{ab} \eta (1-\eta)$ and $\kappa_{\alpha\beta} \approx \chi_{ab}$. Hence the interfacial energy can be estimated as
\begin{equation}
\label{eqn:sigma-chi}
\gamma_{\alpha \beta} \approx \frac{\pi c_0\lambda RT}{4}\chi_{ab}, 
\end{equation}
which is directly proportional to the Flory interaction parameter $\chi_{ab}$. The relation above is used in  Sec.~\ref{sec:design}, where we comment on how the desired packing morphology of coexisting structures can be obtained by appropriately choosing the relations between surface tensions $\lbrace \gamma_{\alpha\beta}\rbrace $, which are functions of the Flory interaction parameters $\lbrace \chi_{ij}\rbrace$.

Note that the expression for the interfacial energy in Eq.~(\ref{eqn:sigma-cal}) assumes that the interface is straight in composition space. However, in our simulations we observed that the interfaces are typically curved in composition space (see, e.g., Fig.~\ref{fig:case2}). Hence, while the expression in Eq.~(\ref{eqn:sigma-cal}) overestimates the true interfacial energy, it still provides a very useful estimate.

\subsection{Kinetics and numerical implementation}
Rather than numerically minimizing the total free energy in Eq.~(\ref{eq:tot_free_energy}) to obtain the morphology of coexisting phases, we instead focus on the dynamic evolution of the mixture. Since the absolute concentrations $\lbrace c_i \equiv c_0 \phi_i \rbrace$ are fixed in our system, we employ the so-called model B dynamics~\cite{hohenberg1977theory}
\begin{equation}
\label{eqn:governing-c}
\frac{\partial c_i}{\partial t} = \nabla \cdot \left[ \sum_{j} M_{ij} \nabla \left(  \frac{\delta f}{\delta c_j} \right)\right],
\end{equation}
where $M_{ij}$ are the Onsager mobility coefficients and $\delta f/\delta c_j$ denotes a functional derivative of the free energy density. 
Furthermore, we adopt Kramer's model~\cite{kramer1984interdiffusion} to model the fluxes, and write the mobility coefficients as $ M_{ij} = (D c_0/RT) \times \left( \phi_i \delta_{ij} - \phi_i \phi_j \right)$. We also assume that all components have identical diffusion coefficients $D_{ij} \equiv D$.
It should be noted that in Eq.~(\ref{eqn:governing-c}) we have omitted terms accounting for any advective hydrodynamic flow behavior and thermal noise. In this paper we focus on the regions of phase space that undergo diffusion-dominated spinodal decomposition, for which the free energy is locally unstable and thermal fluctuations are unimportant~\cite{rogers1989numerical}. Thermal fluctuations are of course important for the nucleation and growth within the binodal regions\cite{granasy2006phase,porter2009phase}, processes which are not investigated in this paper but are left for future work.  

Now, instead of the absolute concentrations $c_i$, we work with relative compositions $\phi_i$ that evolve via
\begin{equation}
\frac{\partial \phi_i}{\partial t} =  D \nabla \cdot \left[ \phi_i \sum_{j}\left(\delta_{ij} - \phi_j \right)  \nabla \tilde{\mu}_j  \right],
\label{eqn:governing-phi}
\end{equation}
where we introduced the dimensionless chemical potentials
\begin{eqnarray}
\tilde{\mu}_j = \frac{\delta \tilde{f}}{\delta \phi_j}=1 + \ln \phi_j + \sum_{k=1}^{N} \chi_{jk}  (1 + \lambda^2 \nabla^2) \phi_k 
\end{eqnarray}
in terms of the dimensionless free energy density $\tilde{f} = f/(c_0RT)$. 

The nonlinear partial differential equations in Eqn.~(\ref{eqn:governing-phi}) were solved numerically in a 3D cubic box with linear dimension $L$ discretized with $128 \times 128 \times 128$ uniform grid points and periodic boundary conditions.
A semi-implicit time-integration scheme \cite{zhu1999coarsening} was used, which enabled us to use relatively large time steps.  
To do so, we first discretized Eqn.~(\ref{eqn:governing-phi}) in time and separated the implicit linear and the explicit non-linear terms following the usual IMEX (implicit-explicit) scheme\cite{ascher1997implicit} as 
\begin{equation}
\label{eqn:governing-numerical}
\frac{\phi_i^{n+1} - \phi_i^{n}}{\Delta t} = N_i(\phi_i^n) + L_i(\phi_i^{n+1}), 
\end{equation}
where $\phi^{n}_i$ is the volume fraction field of component $i$ at time step $n$. $N_i$ and $L_i$ denote the nonlinear and linear parts of the right hand side of Eqn.~(\ref{eqn:governing-phi}), respectively. In the present problem, the stiffest term of the numerical integration corresponds to the $\nabla^4$ operator, which is actually nonlinear, because the mobilities  $\lbrace M_{ij}\rbrace $ are functions of the compositions $\lbrace \phi_i \rbrace $. To overcome this difficulty, we followed the procedure in Ref.~\cite{zhu1999coarsening} and introduced an artificial linear $\nabla^4$ term to stabilize the nonlinear term as 
\begin{eqnarray}
&& N_i(\{\phi_i\}) =  D \nabla \cdot \left[ \phi_i \sum_{j}\left(\delta_{ij} - \phi_j \right) \nabla \tilde{\mu}_j  \right] + A D \lambda^2 \nabla^4 \phi_i,  \nonumber \\
&& L_i(\phi_i) =  - A D\lambda^2 \nabla^4 \phi_i, 
\end{eqnarray}
where the numerical prefactor $A$ is chosen empirically to ensure numerical stability. In Ref.~\cite{zhu1999coarsening} the value $A=0.5\chi_{12}$ was used for the study of binary mixtures. Based on their experience, the value $A=0.5\max \{\chi_{ij}\}$ was used in the present work.

Now, when evaluating nonlinear terms $N_i(\{\phi_i\})$, the products of composition fields $\phi_i^n (\vec r)$ are carried out in real space, while the spatial derivatives are evaluated in Fourier representation $\hat{\phi}_{i}^{n} (\vec k)=\int_V  d \vec r \, e^{-i \vec k \cdot \vec r} \phi_i^n (\vec r)/V$. The Fast Fourier Transform (FFT) algorithm was used to convert back and forth between real space and Fourier space representations~\cite{cooley1969fast}. In Fourier space, the implicit equation~(\ref{eqn:governing-numerical}) can be solved to obtain
\begin{equation}
 \hat{\phi}_{i}^{n+1} = \frac{ \hat{\phi_i^n} + \hat{N}_i(\phi^n_i)\Delta t }{1 + A\lambda^2k^4 D\Delta t},
\end{equation}
where $\hat{\cdot}$ denotes a Fourier transform and $k = |\vec{k}|$ is the magnitude of the wave vector $\vec{k}$.

To make equations dimensionless, the lengths are measured in units of the cubic box size $L$ and time is measured in the units of $\tau=\lambda^2/D$, which describes the characteristic time of diffusion across the interface between two phases. We chose $\lambda/L = 0.45 \times 10^{-2}$ and the time step $\Delta t = \tau/2$. Our system is initialized with the desired average composition $\lbrace \bar{\phi}_i \rbrace$ with some small perturbation (uniform random noise with small magnitude), and then the simulation runs for a total duration of $10^{5}-10^6 \tau$. 

\subsection{Post-processing methods}
The model B dynamics described above can be used to study phase separation of mixtures with an arbitrary number of components $N$. Once the mixture phase separates, we need to find a way to extract the number $P$ of coexisting phases and the compositions $\lbrace \phi_i^\alpha \rbrace$ of each phase. In order to do this, it is useful to represent a simulation configuration in the composition space, where the compositions $\lbrace \phi_i \rbrace$ at each of the $128\times128\times128$ grid points are represented as points in an $(N-1)$-dimensional composition space (see Fig.~\ref{fig:PCA}). Note that there are only $N-1$ independent components due to the constraint $\sum_i \phi_i=1$. Once a mixture phase separates into $P$ coexisting phases, all the data points lie on a $(P-1)$-dimensional manifold in the composition space (see Fig.~\ref{fig:PCA}). Majority of the points are located in $P$ corners that correspond to the compositions $\{\phi_i^\alpha\}$ of $P$ distinct phases denoted with Greek labels. Points that connect these corners correspond to the compositions associated with interfacial regions between phases. Below we describe how this information can be extracted with Principal Component Analysis (PCA) methods to estimate the number $P$ of coexisting phases together with a K-Means clustering algorithm to estimate the compositions of phases $\lbrace \phi_i^\alpha \rbrace$.

\subsubsection{Estimation of the number of coexisting phases with the Principal Component Analysis}
\begin{figure}[!t]
\centering
  \includegraphics[width=0.47\textwidth]{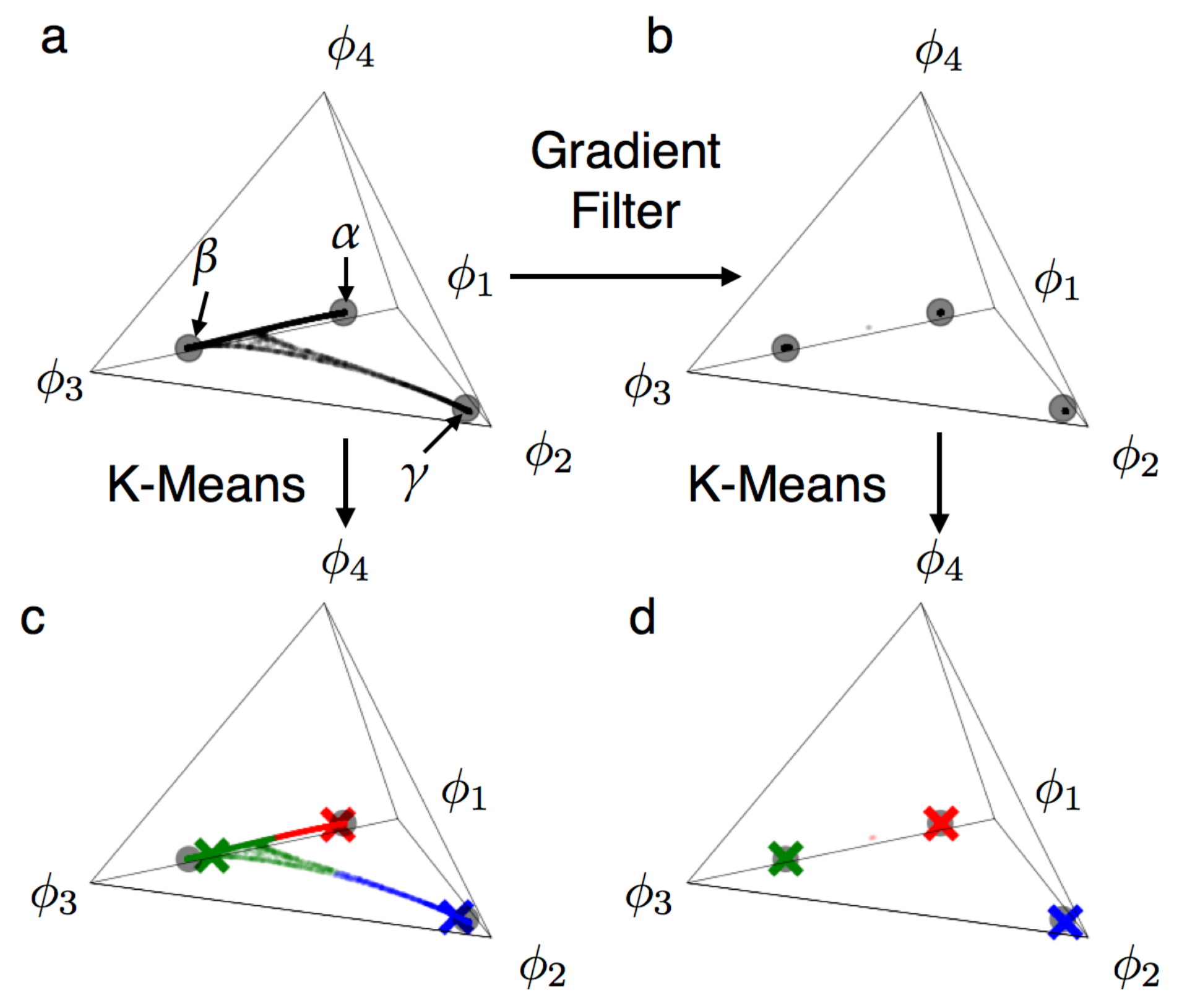}
  \caption{An example of the post-processing procedure for a $N=4$ component mixture with $P=3$ coexisting phases. (a) The composition map of the simulation data. Large gray dots correspond to the equilibrium compositions of the three coexisting phases $\alpha$, $\beta$, $\gamma$ as determined from the convex hull algorithm. (b) The composition map for the filtered simulation data (see text). Note that most points are concentrated in the neighborhood of equilibrium phase compositions. (c-d) K-Means clustering based on the (c) original and (d) filtered simulation data. Crosses mark the cluster centers and data points are colored according to the cluster to which they belong. Data in this figure was obtained from simulations with Flory interaction parameters $\chi_{12} = 4.50,\ \chi_{13} = 2.50, \chi_{23} = 3.50, \chi_{i4} = 1.50,\ (i=1,2,3)$ and initial compositions  $\lbrace \bar{\phi}_i\rbrace = \lbrace 0.30, 0.20, 0.45, 0.05\rbrace$. }
  \label{fig:PCA}
\end{figure}

The PCA method can be thought of as the fitting of an $N$-dimensional ellipsoid to the composition data, where each axis of the ellipsoid represents a principal component~\cite{jolliffe2011principal}. As can be seen in Fig.~\ref{fig:PCA} the composition points lie approximately on a $(P-1)$-dimensional hyperplane. Therefore, the PCA method produces an ellipsoid with $P-1$ axes of finite size, while the other $N-P+1$ axes are nearly zero.

First, we construct the dataset ${\bf X}$ for the PCA. The composition $\lbrace \phi_{i} \rbrace$ for each of the $128\times128\times128$ grid points is treated as one entry in the dataset ${\bf X}$, which is thus a matrix of dimensions $128^3\times N$. Second, we construct the covariance matrix ${\bf C} = {\bf X}^{T} {\bf X}$ of dimension $N\times N$ and calculate its eigenvalues and eigenvectors. Eigenvectors and eigenvalues in turn correspond to the directions and lengths of ellipsoid axes, respectively. For the solution with $P$ coexisting phases, we thus expect $P-1$ non-zero eigenvalues and $N-P+1$ eigenvalues that are nearly zero. However, as can be seen in Fig.~\ref{fig:PCA}, the interfacial points that connect certain two phases do not necessarily lie on a straight line. Due to the curvature of interfaces in the composition space some points may reisde outside the $(P-1)$-dimensional hyperplane, and in such cases, the PCA analysis produces additional nonzero eigenvalues. This problem can be avoided by removing the interfacial points, which correspond to regions with large compositional gradients $\Delta  = \max_{i} |\nabla\phi_i|^2$. By filtering out interfacial points with gradients larger than $\Delta= 0.002/\lambda^2$, (see  Fig.~\ref{fig:PCA}), we kept only points that correspond to $P$ bulk phases. After the filtering, the PCA analysis in fact produces only $P-1$ nonzero eigenvalues. For the $N=4$ component mixture with $P=3$ coexisting phases in Fig.~\ref{fig:PCA}, the eigenvalues of the covariance matrix ${\bf C}$ are $1.69\times 10^{-1}$, $6.51\times 10^{-2}$, $1.01\times 10^{-5}$, and $2.34\times 10^{-12}$. In practice, we find that there is good agreement for the number of coexisting phases $P$ with the convex hull algorithm described in the previous section, if we define nonzero eigenvalues as those that are larger than $10^{-4}$.

\subsubsection{Estimation of phase compositions with K-Means}
Compositions of stable phases correspond to regions of densely clustered points in the composition space (see Fig.~\ref{fig:PCA}). Therefore, once we determine the number $P$ of coexisting phases with the PCA method, we can then use the standard K-Means clustering method~\cite{hartigan1979algorithm} to compute the centers of clusters, which yield the compositions of phases $\{\phi_i^\alpha\}$. In the present work, we employed the scikit-learn package~\cite{sklearn_api} to compute the centers of clusters. In analogy with PCA method, it is important to filter out the interfacial points, otherwise the centers of clusters may be shifted away from the actual compositions (see Fig.~\ref{fig:PCA}). With the caveats noted above, the compositions of phases obtained from the K-means clustering method agree very well with the compositions obtained from the convex hull method described in Section 2 (see Fig.~\ref{fig:PCA}). 

Once the compositions of phases are known, we can use this information to construct local phase indicator functions $\lbrace \eta_\alpha(\vec r) \rbrace$ such that 
\begin{equation}
\eta_{\alpha}(\vec r) = 
\begin{cases}
1 \quad \text{in the bulk phase $\alpha$},  \\
0 \quad \text{in the bulk of all other phases} .
\end{cases}
\label{eq:indicator_functions}
\end{equation}
For each grid point $\vec r$, we can prescribe that the local composition $\lbrace \phi_i(\vec r) \rbrace$ is a mixture of $P$ phases with volume fractions $\lbrace \eta_\alpha (\vec r) \rbrace$, such that
\begin{equation}
\label{eq:order_parameters}
\phi_{i}(\vec{r}) = \sum_{\alpha} \phi_i^{\alpha} \eta_{\alpha}.
\end{equation}
These phase indicator functions can be thought of as proxies for the intensity of fluorescent markers that are often employed in experiments to mark individual phases. In the bulk of each phase $\beta$ the local concentration $\phi_{i}(\vec{r})\approx  \phi_i^\beta $ and hence $\eta_\alpha (\vec r) \approx \delta_{\alpha \beta}$. 
The system of $N$ equations for the $P\le N$ unknowns $\lbrace \eta_\alpha (\vec r) \rbrace$ in Eq.~(\ref{eq:order_parameters}) can be approximately solved by calculating the pseudo-inverse of the $N\times P$ matrix ${\Phi}\equiv \phi_i^\alpha$. The phase indicator functions can then be calculated as 
\begin{equation}
\label{eq:order_parameters2}
\eta_{\alpha} = R_{\alpha j} \phi_{j}, \ \ {\bf R} = \left( { \Phi}^T { \Phi}  \right)^{-1} {\Phi}^T.
\end{equation}
Note that within the interfacial regions the values of $\eta_\alpha$ may become negative or larger than 1. To remedy this, we apply the following transformation to regularize the indicator functions~\cite{cogswell2011thermodynamic}: we set $\eta_{\alpha} = 1$ if $\eta_{\alpha} > 1$, and $\eta_\alpha=0$ if $\eta_{\alpha} < 0$. After this, we normalize the indicator functions such that $\sum_{\alpha} \eta_{\alpha} = 1$. In this way, we ensure that $\eta_\alpha \in [0, 1]$ and represents the fractions of different phases at a given location.

\section{Morphology of coexisting phases}
\label{sec:4}
\begin{figure}[!t]
\centering
  \includegraphics[width=0.45\textwidth]{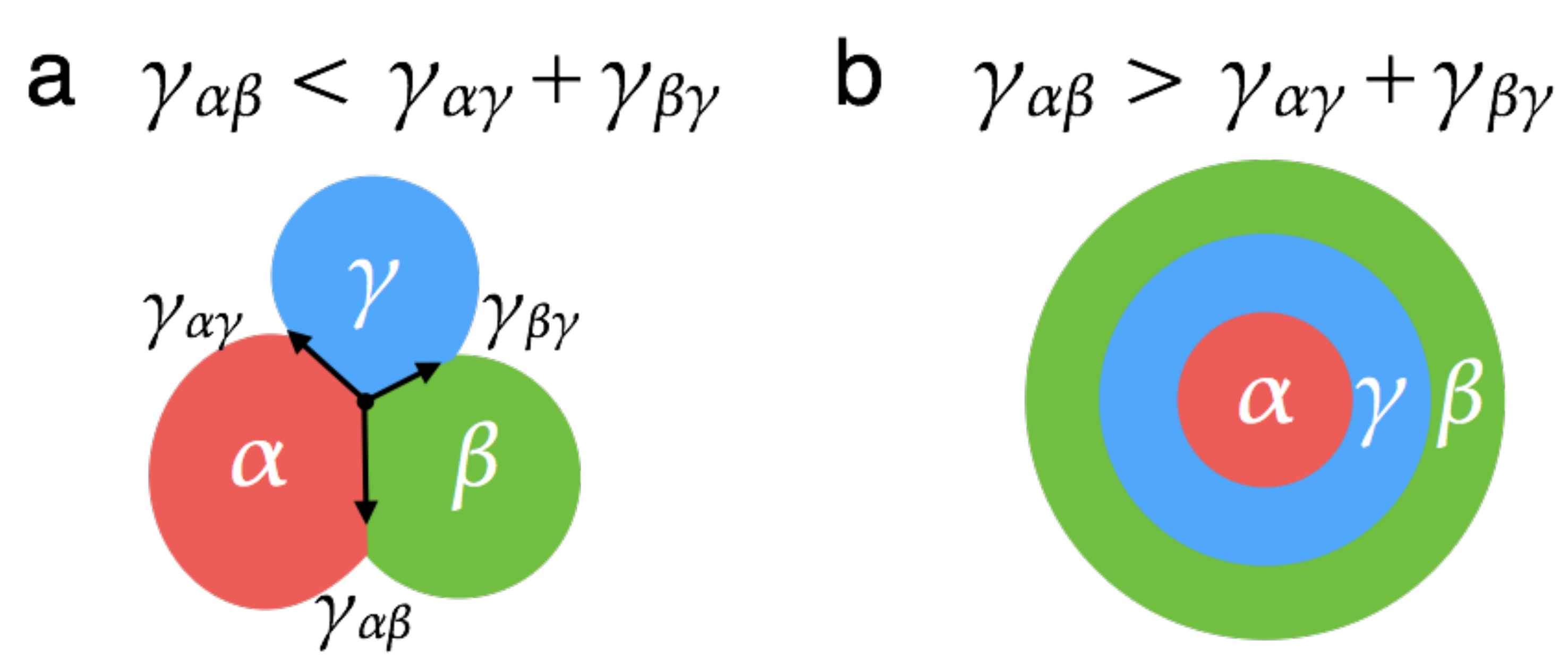}
  \caption{Schematic of morphologies in a system with three phases $\alpha$, $\beta$, and $\gamma$ based on the magnitudes of surface tensions $\gamma_{\alpha\beta}\ge \gamma_{\alpha\gamma} \ge \gamma_{\beta\gamma}$. (a)~Mechanically stable triple-phase junctions. Finite contact angles between different phases are related to the force balance via surface tensions. (b)~Mechanically unstable triple-phase junctions. Phase $\gamma$ completely wets phases $\alpha$ and $\beta$ so as to minimize the overall interfacial energy.}
  \label{fig:surface}
\end{figure}
In previous sections we described how the number of coexisting phases $P$ and their compositions $\lbrace \phi_i^\alpha \rbrace$ can be obtained either with the convex hull construction of the free energy function (Sec.~\ref{sec:phase_diagram}) or by analyzing the dynamic evolution of the mixture together with the PCA and K-Means clustering methods (Sec.~\ref{sec:kinetics}).
In this section we compare the results of these two approaches for the case of 4-component mixtures. Furthermore, we investigate the microstructure of $P$ coexisting phases that depends on both the surface tensions $\lbrace\gamma_{\alpha\beta}\rbrace$ between phases and on the volume fractions of the phases, which are functions of interaction parameters $\lbrace\chi_{ij}\rbrace$ and average compositions $\lbrace \overline{\phi}_i \rbrace$, respectively. Note that for any triplets of phases $\alpha$, $\beta$, and $\gamma$, with surface tensions $\gamma_{\alpha\beta}\ge \gamma_{\alpha\gamma} \ge \gamma_{\beta\gamma}$, the triple-phase junctions are mechanically stable (unstable) when $\gamma_{\alpha\beta} < \gamma_{\alpha \gamma} + \gamma_{\gamma\beta}$ ($\gamma_{\alpha\beta} > \gamma_{\alpha \gamma} + \gamma_{\gamma\beta}$)~\cite{brochard2002capillary}. When triple-phase junctions are mechanically unstable, the phase $\gamma$ penetrates between phases $\alpha$ and $\beta$ to minimize the total interfacial energy (see Fig.~\ref{fig:surface}). The $\binom{P}{3}$ inequalities between surface tensions thus dictate the equilibrium morphology of phase separated mixtures. We show that the packing morphologies found in simulations are consistent with the estimated surface tensions in Eq.~(\ref{eqn:sigma-cal}) from the Cahn-Hilliard formalism. While the microstructure is primarily determined from equilibrium properties, we show an example where kinetic pathways, which lead to multi-stage phase separation, are responsible for the formation of ``pearled-chain'' structures.

\begin{figure*}[!t]
\centering
  \includegraphics[width=0.9\textwidth]{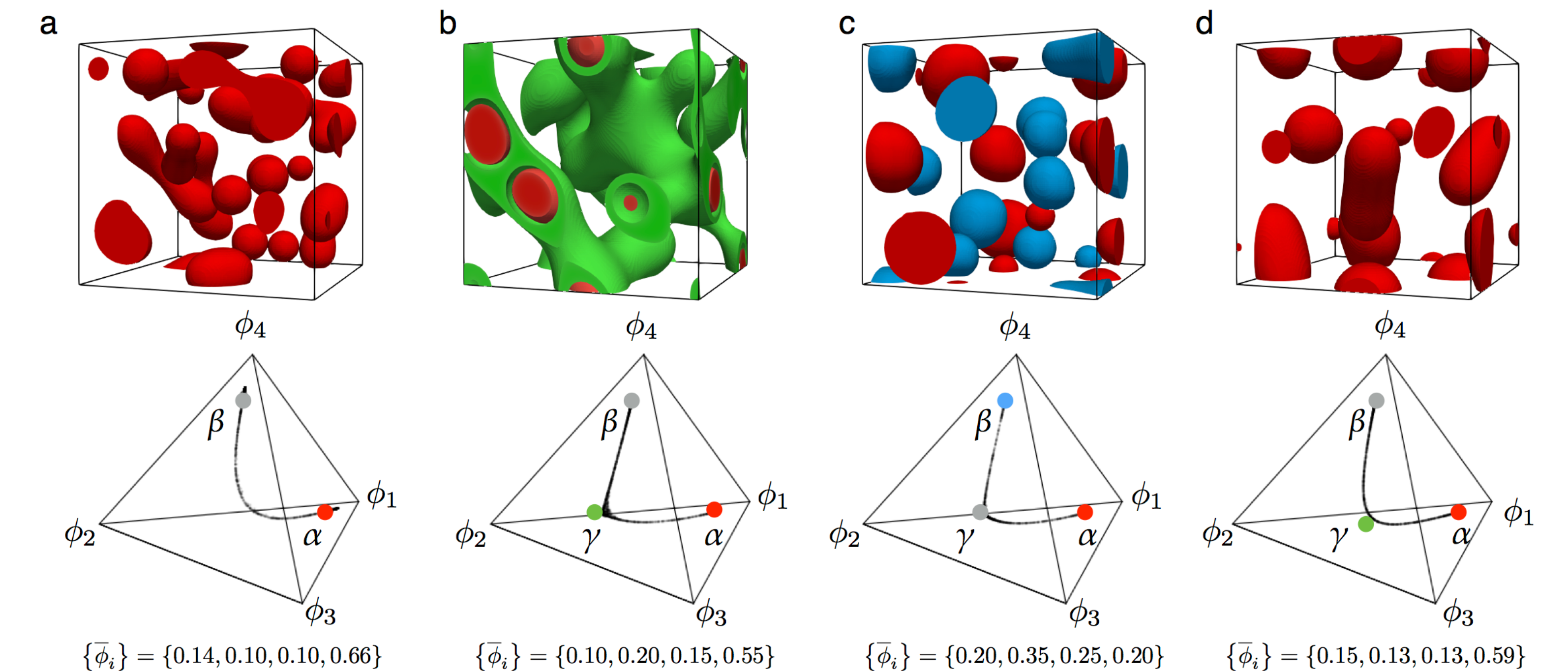}
  \caption{Four distinct morphologies of 4-component mixtures that include a pair of strongly interacting components. (a)~Stable two-phase region, (b-c)~stable three-phase regions, and (d)~metastable two-phase region. Bottom row displays points in the composition space, where large colored dots mark the phase compositions obtained from the convex hull algorithm.  Top row displays indicator functions of phases in real space (colors correspond to the colored dots in the bottom row). The majority phase, which is marked with the gray dot in the composition map, is transparent in top rows. The interaction parameters were set to $\chi_{14} = \chi_{41} = 5.50$ and all others $\chi_{ij} = 2.70, (i\neq j)$.}
  \label{fig:case2}
\end{figure*}

\subsection{Symmetric quench}
First, we analyze the simplest possible case, where all the interaction parameters are equal $\chi_{ij} \equiv \chi$, $(i\neq j)$, and also the average compositions for all components are equal to $\bar{\phi}_i \equiv 1/N$. The $N$-component mixture is thus completely symmetric and either stays mixed in a single phase, or phase separates into $N$ coexisting phases, each of which is enriched with one of the components. In each phase, the $N-1$ minority components have a composition $0<\phi\leq(1/N)$, while the majority component has composition $1 - (N-1)\phi$. Note that the $\phi=1/N$ case corresponds to an initially equimolar mixture. Due to the symmetry of the system, the free energy density can be expressed in terms of a single variable $\phi$ as
\begin{eqnarray}
\label{eqn:f-symmetric}
&&\tilde f_{FH}(\phi) = (N-1) \phi \ln \phi + \left[1 - (N-1)\phi\right]\ln\left[1 - (N-1)\phi\right]   \nonumber \\
&& \quad \quad \ \ \ \ \, \, + \ \chi (N-1)\phi \left(1 - \frac{N}{2} \phi \right),
\end{eqnarray}
 where $\tilde f_{FH}(\phi)= f_{FH}(\phi)/(c_0 RT)$ is the dimensionless free energy density.
 
This special case can thus be mapped to an equivalent binary mixture, which can be analyzed with standard tools. For sufficiently large value of the interaction parameter $\chi > \chi_c$, the system phase separates into $N$ coexisting phases. For a symmetric solution with many components ($N\gg 1$) we find that the critical interaction parameter scales as $\chi_c \sim 2 \ln N$ and in each phase the concentration of minority components scales as $\phi \sim 1/N^2$. Note that the spinodal region, where the free energy becomes locally unstable for the uniform mixture, is achieved only when the interaction parameter becomes very large $(\chi \ge N \gg \chi_c)$. Thus, for $\chi \approx \chi_c$, the system phase separates via nucleation and growth by crossing an energy barrier $\Delta \tilde f \sim (\ln N)/4$, as estimated from Eq.~(\ref{eqn:f-symmetric}).

\subsection{Pair of strongly repelling components}
\label{sec:pair_strong_int}
Next, we investigate a slightly more complicated case, where a pair of two components repel very strongly (large value of $\chi)$, while all the other interactions are moderate. As a representative system, we take a 4-component solution, where the components $1$ and $4$ interact strongly ($\chi_{14}=5.50$), while for all other components, $\chi_{ij}=2.70$. Because of the strong repulsion, the system typically phase separates into at least two phases (see Fig.~\ref{fig:case2}), where one of the phases ($\alpha$) is enriched with component $1$ and devoid of component $4$, while another phase ($\beta$) is enriched with component $4$ and devoid of component $1$. Note that when the average composition $\lbrace \overline \phi_i \rbrace$ is in a region of composition space, where the system separates into $3$ phases, then the additional phase $\gamma$, which is enriched with components $2$ and $3$, penetrates between the phases $\alpha$ and $\beta$ in order to minimize the total interfacial energy (see Fig.~\ref{fig:case2}b,c). This happens whenever surface tensions satisfy the inequality
\begin{equation}
\label{eqn:sigma-triplet}
\gamma_{\alpha \beta} > \gamma_{\alpha \gamma} + \gamma_{\beta \gamma},
\end{equation}
which makes the triple-phase junctions mechanically unstable. For the parameters used in Fig.~\ref{fig:case2}b,c, we estimated dimensionless surface tensions $\lbrace \tilde{\gamma}_{\alpha \beta} = 0.708,\ \tilde{\gamma}_{\alpha \gamma} = 0.109, \, \tilde{\gamma}_{\beta \gamma} = 0.107 \rbrace$ from the Cahn-Hilliard procedure in Eqn.~(\ref{eqn:sigma-cal}), where dimensionless surface tensions are defined as $\lbrace \tilde \gamma_{\alpha\beta}\rbrace \equiv \lbrace \gamma_{\alpha\beta} / (2\lambda c_0 R T)\rbrace$. The estimated surface tensions are consistent with the inequality in Eqn.~(\ref{eqn:sigma-triplet}). While the phase $\gamma$ penetrates between phases $\alpha$ and $\beta$ in both Fig.~\ref{fig:case2}b and \ref{fig:case2}c, the two morphologies are quite distinct due to differences in the volume fractions of the three phases. 

Here, we make another observation. In some regions of composition space, the convex hull construction predicts $3$ coexisting phases, while in simulations of model B dynamics we see only $2$ coexisting phases (see Fig.~\ref{fig:case2}d). This is due to the fact that our dynamics is restricted to the spinodal region, and hence the system can get trapped in metastable states in the absence of thermal noise.

\subsection{Multistage phase separation}
\begin{figure*}[!ht]
\centering
  \includegraphics[width=0.8\textwidth]{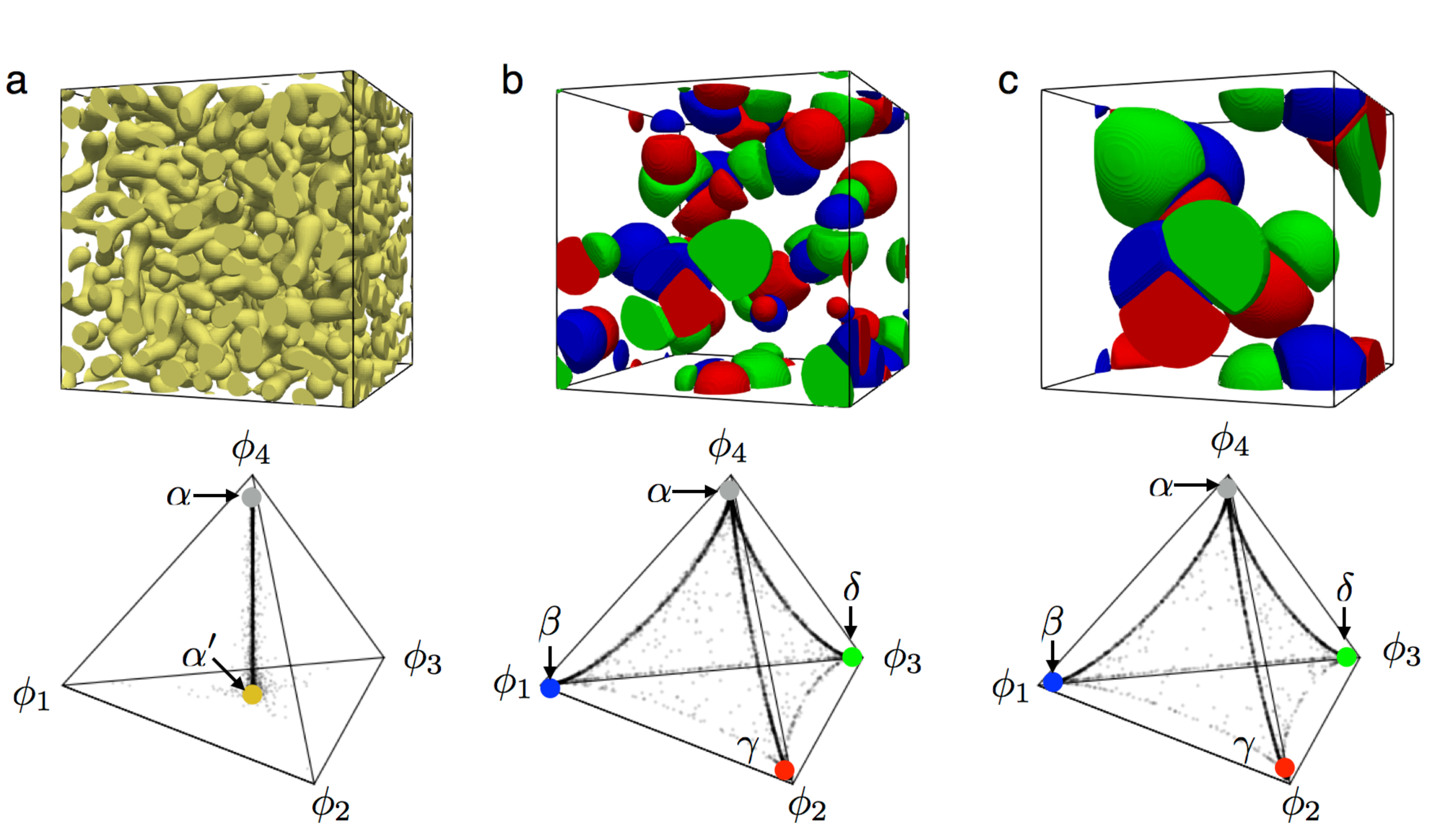}
  \caption{Multistage phase separation. a) At $t = 1,000 \, \tau$, the initial instability leads to the formation of two phases $\alpha$ and $\alpha'$. b) At $t=150,000 \,\tau$, a secondary instability causes phase $\alpha'$ to split into three equilibrium phases $\beta$, $\gamma$, and $\delta$, which form the ``pearl chain''--like structure. c)  At $t=292,000\, \tau$, pearled chains break into triplet ``Janus particle''-like droplets due to the Plateau--Rayleigh instability. Bottom row displays points in the composition space, where large colored dots mark the phase compositions obtained from the convex hull algorithm.  Top row displays indicator functions of phases in real space (colors correspond to the colored dots in the bottom row). The majority phase, which is marked with the gray dot in the composition map, is transparent in top rows. Interaction parameters were set to $\chi_{ij} \equiv 4.50, (i\neq j)$, with an average composition $\lbrace \bar{\phi}_i\rbrace = \lbrace 0.1, 0.1, 0.1, 0.7 \rbrace$.}
  \label{fig:case3}
\end{figure*}

In binary mixtures, spinodal decomposition occurs instantaneously, while in mixtures with more components, phase separation can happen in several stages. Here, we report on an example of such behavior in a 4-component mixture with symmetric interaction parameters $\chi_{ij} \equiv 4.5, \ (i\neq j)$. The solution with average composition $\lbrace  \bar{\phi}_i \rbrace = \lbrace 0.10, 0.10, 0.10, 0.70\rbrace$ first separates into 2 phases, and subsequently one of the phases demixes into 3 phases (see Fig.~\ref{fig:case3} and \href{http://www.princeton.edu/~akosmrlj/videos/}{Video~S2}).

This can be understood by considering the local stability of the free energy function. The Hessian matrix $H_{ij} = \frac{\partial^2 f_{FH}}{\partial\phi_i \partial \phi_j}$ evaluated at the initial composition, has one negative value with the corresponding eigenvector $\lbrace \phi_i^e \rbrace =\lbrace -0.2, -0.2, -0.2,\ 0.94 \rbrace$. At early stages of the phase separation process, the mixture thus initially forms two phases by following the free energy gradients, which are initially primarily oriented in the direction of the eigenvector $\lbrace \phi_i^e \rbrace$. By minimizing the free energy in the direction of the eigenvector $\lbrace \phi_i^e \rbrace $ we find two local minima located at $\lbrace \phi_i^{\alpha} \rbrace = \lbrace 0.0234,\ 0.0234,\ 0.0234,\ 0.9298\rbrace$ and $\lbrace \phi_i^{\alpha'} \rbrace = \lbrace 0.31,\ 0.31,\ 0.31,\ 0.07\rbrace$. These are approximately the compositions of the two phases $\alpha$ and $\alpha'$ that form at the initial stages of the phase separation process (see Fig.~\ref{fig:case3}a). By analyzing the eigenvalues of Hessian matrix at compositions $\lbrace \phi_i^{\alpha} \rbrace$ and $\lbrace \phi_i^{\alpha'} \rbrace$ we find that the phase $\alpha$ corresponds to a local minimum (positive eigenvalues), while the phase $\alpha'$ actually corresponds to a saddle point (two negative eigenvalues). Therefore the phase $\alpha'$ eventually phase separates into 3 phases (see Fig.~\ref{fig:case3}b,c), leading to the emergence of all 4 equilibrium phases.

We can also rationalize the resulting morphology of the system via the following arguments. From Eq.~(\ref{eq:volume_fractions}) we can estimate the volume fractions $\eta_\alpha=0.7$ and $\eta_{\alpha'}=0.3$ of the two phases that form initially. Because the volume fraction of phase $\alpha'$ is above the percolation threshold~\cite{stauffer2014introduction}, the two phases form a bicontinuous structure. After $\alpha'$ phase separates, the three new phases form within a bicontinuous structure. As a consequence, the system initially forms ``pearl chain''--like structures of the $3$ phases (see Fig.~\ref{fig:case3}b), while during the later coarsening stage, these long chains break into droplets, courtesy of the Plateau--Rayleigh instability~\cite{eggers1997nonlinear,brochard2002capillary}, leading to the formation of triplet ``Janus-like'' droplets. We note that if the volume fraction of the intermediate phase $\alpha'$ was below the percolation threshold, then the system would first form droplets of the phase $\alpha'$, which subsequently phase separate into triplet ``Janus particle''-like droplets.

\subsection{Nested structures}

\begin{figure*}[!t]
\centering
  \includegraphics[width=0.8\textwidth]{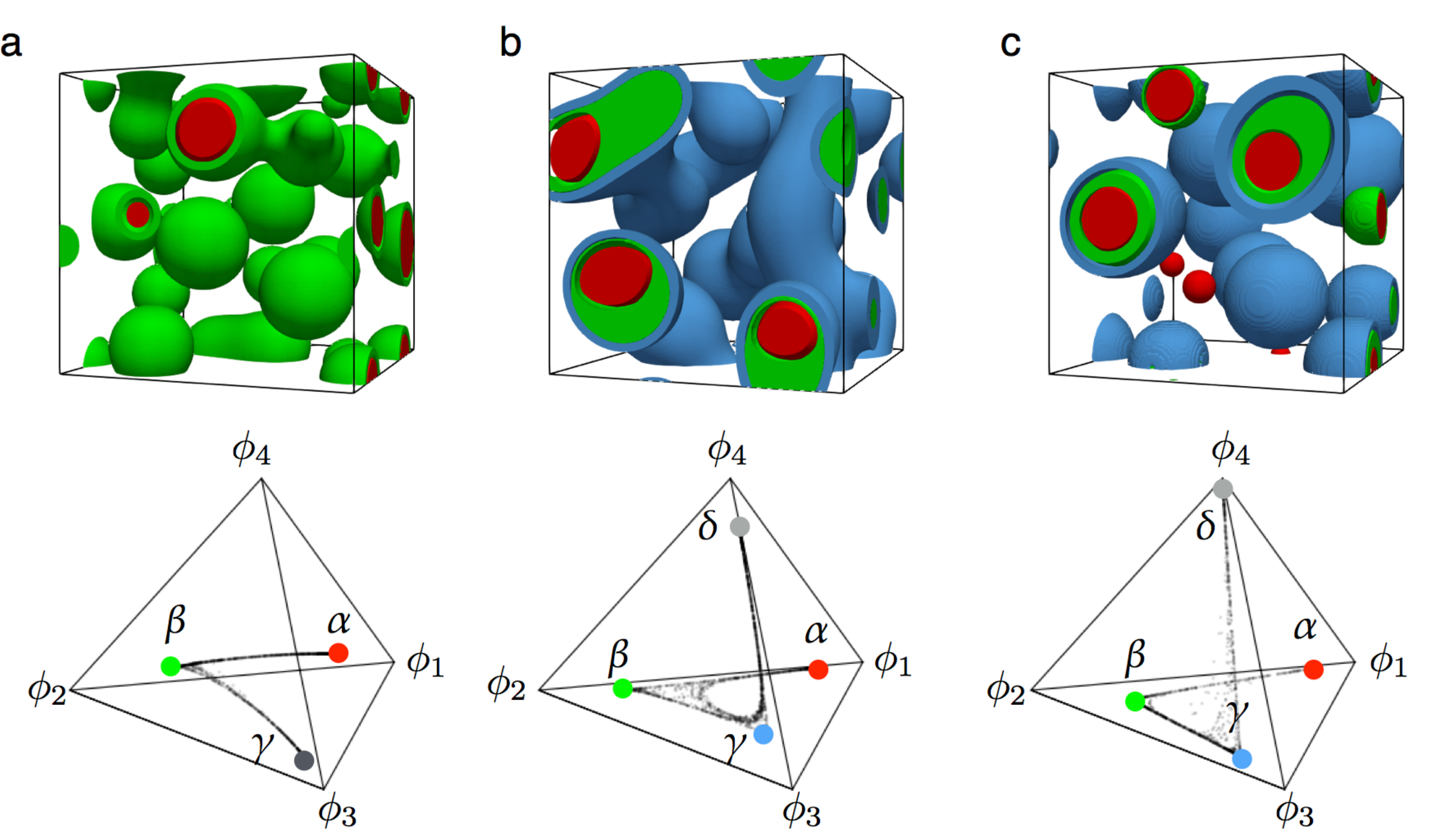}
  \caption{Nested ``Russian doll'' droplets. (a) A $3$-phase ``Russian doll'', (b) an improper $4$-phase ``Russian doll'', and (c) a proper $4$-phase ``Russian doll''. Bottom row displays points in the composition space, where large colored dots mark the phase compositions obtained from the convex hull algorithm.  Top row displays indicator functions of phases in real space (colors correspond to the colored dots in the bottom row). The majority phase, which is marked with the gray dot in the composition map, is transparent in top rows.
The interaction parameters and initial compositions were set to:
(a) $\chi_{12} = 2.50,\ \chi_{13} = 5.50,\ \chi_{23} = 3.50,\ \chi_{i4} = 1.50,\,(i = 1,2,3)$
and $\lbrace  \bar{\phi}_i \rbrace = \lbrace 0.10, 0.15, 0.70, 0.05\rbrace$;
(b) $\chi_{12} = 2.50,\ \chi_{13} = 4.00,\ \chi_{23} = 3.00,\ \chi_{14} = 5.50, \ \chi_{24} = 5.00, \ \chi_{34} = 2.50$
and $\lbrace  \bar{\phi}_i \rbrace = \lbrace 0.10, 0.10, 0.20, 0.60\rbrace$;
(c) $\chi_{12} = 2.50,\ \chi_{13} = 4.10,\ \chi_{23} = 2.40,\ \chi_{14} = 7.00, \ \chi_{24} = 5.10, \ \chi_{34} = 3.70$,
and $\lbrace  \bar{\phi}_i \rbrace = \lbrace 0.06, 0.09, 0.12, 0.73\rbrace$.
}
  \label{fig:case4}
\end{figure*}

In this section we briefly comment on how one can rationally design nested ``Russian doll''-like droplets by tuning the surface tensions between different phases. In Sec.~\ref{sec:pair_strong_int} we already mentioned that in order to make a nested structure with 3 phases $\alpha$, $\beta$ and $\gamma$, surface tensions have to satisfy the inequality in Eq.~(\ref{eqn:sigma-triplet}). The nested structure in Fig.~\ref{fig:case2}b satisfies this inequality, but does not form droplets, as the volume fraction of the intermediate green phase is large enough that it percolates. However, once the inequality between surface tensions is satisfied, then it is straightforward to tune the volume fractions of the coexisting phases by changing the average compositions $\lbrace\overline \phi_i \rbrace$ [see Eq.~(\ref{eq:volume_fractions})] to get the ``core-shell'' droplets for the 3 phases structures (see Fig.~\ref{fig:case4}a).

Next, we will design a morphology with nested ``Russian-doll'' droplets with $4$ coexisting phases, such that phase $\alpha$ domains reside completely inside phase $\beta$ domains, which themselves reside inside phase $\gamma$ domains, which in turn are surrounded by  domains corresponding to phase $\delta$. To ensure that the triple-phase junctions between any of the possible $\binom{4}{3}=4$ triplets are mechanically unstable, we now have 4 different inequalities for surface tensions
\begin{eqnarray}
\label{eq:surface_tension_inequalities}
&& \gamma_{\alpha \gamma} > \gamma_{\alpha \beta} +  \gamma_{\beta \gamma}, \quad \gamma_{\alpha \delta} > \gamma_{\alpha \beta} +  \gamma_{\beta \delta}, \nonumber \\
&& \gamma_{\beta \delta} > \gamma_{\beta \gamma} +  \gamma_{\gamma \delta}, \quad \gamma_{\alpha \delta} > \gamma_{\alpha \gamma} +  \gamma_{\gamma \delta}.
\end{eqnarray}
Note that if any of the above inequalities is not satisfied, then some triple-phase-junctions are mechanically stable and the desired nested structure does not form. An example of such behavior is displayed in  Fig.~\ref{fig:case4}b, where the inequality for the triplet $\alpha,\ \beta,\ \delta$ is slightly violated based on the estimated surface tensions $\lbrace \tilde{\gamma}_{\alpha \beta} = 0.090,\ \tilde{\gamma}_{\alpha \gamma} = 0.474,\ \tilde{\gamma}_{\alpha \delta} = 0.881, \
\tilde{\gamma}_{\beta \gamma} = 0.264, \ 
\tilde{\gamma}_{\beta \delta}=0.860, \ 
\tilde{\gamma}_{\gamma \delta} = 0.142 
\rbrace$, and, as a consequence, the red phase $\alpha$ appears ``pinned'' to the boundary with other phases.

The final question that remains is how do we choose Flory interaction parameters $\lbrace \chi_{ij}\rbrace$, such that the surface tension inequalities in Eq.~(\ref{eq:surface_tension_inequalities}) are all satisfied? We note that in the 4-component mixture, the 4 coexisting phase regions typically form only, when the interaction parameters $\lbrace \chi_{ij}\rbrace$ are quite large. In this case, each phase is enriched with one of the components, and thus the surface tensions between different phases are approximately proportional to the interaction parameters [see Eqn.~(\ref{eqn:sigma-chi})]. Therefore, we can translate the inequalities for surface tensions in Eq.~(\ref{eq:surface_tension_inequalities}) to similar inequalities for interaction parameters $\lbrace \chi_{ij}\rbrace$. Using this idea we were able to construct nested ``Russian-doll'' droplets with 4 coexisting phases (see Fig.~\ref{fig:case4}c and \href{http://www.princeton.edu/~akosmrlj/videos/}{Video~S3}). We verified that the estimated surface tensions 
$ \lbrace \tilde{\gamma}_{\alpha \beta} = 0.113,\ \tilde{\gamma}_{\alpha \gamma} = 0.457,\ \tilde{\gamma}_{\alpha \delta} = 1.64, \
\tilde{\gamma}_{\beta \gamma} = 0.0752, \ 
\tilde{\gamma}_{\beta \delta}=0.940, \ 
\tilde{\gamma}_{\gamma \delta} = 0.595 \rbrace$ 
satisfy the inequalities in Eq.~(\ref{eq:surface_tension_inequalities}).

\section{Domain coarsening kinetics}
\label{sec:5}

Next, we turn to the quantitative description of domain growth and coarsening kinetics during phase separation of multicomponent mixtures. Upon quenching into the spinodal regions of the phase diagram, small compositional heterogeneities are rapidly amplified in time and lead to the formation of compositional domains with a characteristic length scale, as evident in Fig.~\ref{fig:case3}a. Once the spinodal instability is exhausted, phase separating systems enter a so-called domain coarsening regime, during which capillary forces drive competitive growth of larger domains at the expense of smaller ones so as to minimize the total interfacial energy of the system. In two phase liquid systems, coarsening can be quantitatively captured via a single time-dependent length scale (average domain size of the minority phase droplets) $R(t) \sim t^a$, where the coarsening exponent $a=1/3$ for systems in which diffusive transport processes dominates over advective ones \cite{bray2002theory,bray1989exact, berry2018physical,lifshitz1961kinetics}. Importantly, in this so-called scaling regime, morphologies are self-similar, and structural correlation functions only depend on $r/R(t)$. Bray~\cite{bray1989exact,bray2002theory} has argued that when scaling behavior is observed in systems with more than two coexisting phases, the coarsening exponent should still be given by $a=1/3$. Below, we first introduce appropriate structure factors and subsequently examine domain coarsening kinetics in systems with $4$ coexisting phases in light of Bray's theoretical prediction.   


\begin{figure}[!t]
\centering
\includegraphics[width=0.48\textwidth]{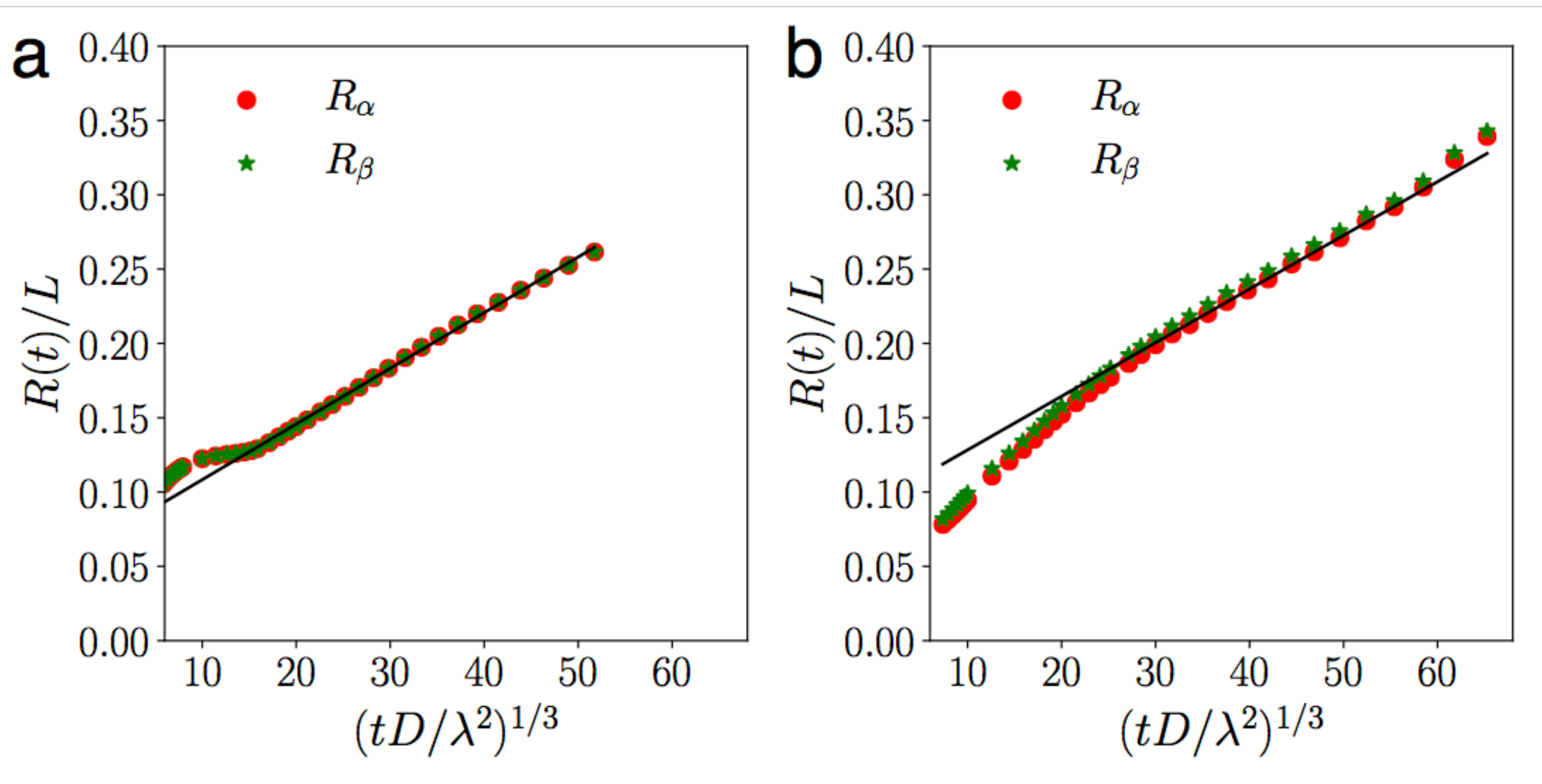}
  \caption{Coarsening kinetics of two coexisting phases for (a) a  binary mixture (N=2) with an interaction parameter $\chi_{12}=2.50$ and composition $\lbrace \overline \phi_i \rbrace = \lbrace 0.5, 0.5 \rbrace$, and  (b) the 4-component mixture in Fig.~\ref{fig:case2}d that is trapped in a metastable state. Solid black lines are linear fits to $R_\alpha$ at large times.}
\label{fig:coarsen-2}
\end{figure}

As mentioned already in Section~\ref{sec:kinetics}, we employ a family of phase indicator functions $\lbrace \eta_{\alpha} \rbrace$ to characterize the morphology of the phase separating $N$-component system. The indicator functions are constructed such that, within domains of a particular phase $\beta$, $\eta_{\beta} = 1$, while all other $\eta_{\alpha} = 0$. In order to quantitatively calculate the characteristic length scale of domains belonging to a specific phase, we introduce the following structure factors
\begin{eqnarray}
&&S_{\alpha \beta}\big(\, \vec{k}, t\big) =    \hat{\eta}_{\alpha}\big(\,\vec{k}, t\big)\ \hat{\eta}_{\beta}\big(-\vec{k}, t\big), 
\end{eqnarray} 
where $\hat{\eta}_{\alpha} \big(\,\vec k,t\big)$ denote the Fourier transforms of phase indicator functions defined as $\hat{\eta}_{\alpha} \big(\,\vec k,t\big)=\int_V  d^3 \vec r \, e^{-i \vec k \cdot \vec r} \eta_\alpha (\,\vec r,t)/V$.
Given a structure factor $S_{\alpha \beta}\big(\, \vec{k}, t\big)$, we define our characteristic length scale in a commonly adopted way~\cite{koga1991spinodal,furukawa2000spinodal,kendon2001inertial,camley2010dynamic} via
\begin{equation}
R_{\alpha}(t) = 2\pi \frac{\int \! d^3\vec{k} \, S_{\alpha\alpha}\big(\,\vec{k}, t\big ) }{\int\! d^3\vec{k}\, \big|\vec{k}\big|\, S_{\alpha\alpha}\big(\,\vec{k}, t\big)  }.
\end{equation}

As a benchmark, we first analyzed the coarsening of a binary mixture, which reaches the usual asymptotic coarsening behavior with exponent $a=1/3$ at around $(tD/\lambda^2)^{1/3} \approx 15$ (see Fig.~\ref{fig:coarsen-2}a). We first compare it to the coarsening of a 4-component mixture that is trapped in a metastable region with two coexisting phases (see Fig.~\ref{fig:case2}d). This mixture also reaches the $t^{1/3}$ asymptotic coarsening stage, but at a later time $(tD/\lambda^2)^{1/3} \approx 25$ (see Fig.~\ref{fig:coarsen-2}b). Small deviations at large times can be attributed to the presence of finite size effects.

\begin{figure}[t]
\centering
\includegraphics[width=0.48\textwidth]{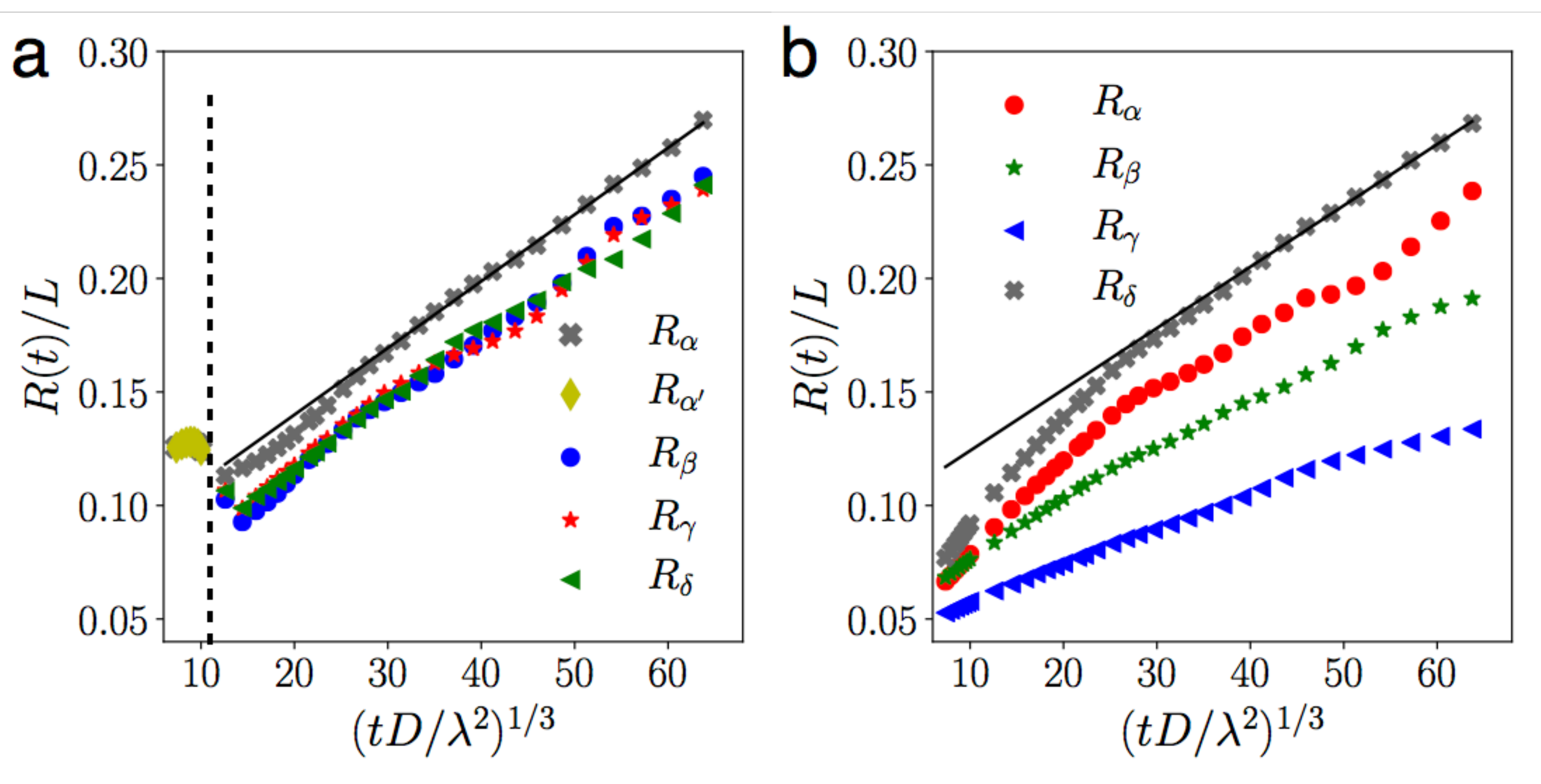}
  \caption{Coarsening of 4-component mixtures with 4 coexisting phases. (a) Coarsening kinetics of the mixture with multistage phase separation displayed in Fig.~\ref{fig:case3}. The transition from the initial instability to the secondary one is denoted by the dashed line. (b) Coarsening kinetics for the mixture with nested ``Russian-doll'' droplet morphology shown in Fig.~\ref{fig:case4}c. Solid black lines are linear fits for the characteristic length scale of the majority phase (marked with gray crosses) at large times.}
\label{fig:coarsen-4}
\end{figure}

In the case of 4-component mixtures with 4 coexisting phases, our numerical coarsening data indeed indicate convergence towards the predicted $t^{1/3}$ behavior~\cite{bray1989exact,bray2002theory}, as shown in Fig.~\ref{fig:coarsen-4}. Very little coarsening takes place during the first stage of the multistage phase separation process involving two coexisting phases displayed in Fig.~\ref{fig:case3}. Once all 4 coexisting phases have emerged, however, the domains of all phases begin to coarsen at the same rate and the asymptotic coarsening is achieved at around $(tD/\lambda^2)^{1/3} \approx 30$. On the other hand, in the case of the ``Russian doll'' morphology in Fig.~\ref{fig:case4}c, all phases appear roughly simultaneously, and begin to coarsen, albeit with different rates. We attribute this to the fact that the initial formation of the nested microstructure imposes correlations on the local compositions, which are not accounted for in standard coarsening theories. On the other hand, once the nested microstructure has formed, a single length scale is sufficient to describe the morphology, and hence a crossover to the predicted $t^{1/3}$ behavior is reached at around $(tD/\lambda^2)^{1/3} \approx 30$.  Small deviations at large times are again attributed to the presence of finite size effects. We expect that the crossover time scale depends on the number of coexisting phases and the droplet morphology, and plan to investigate this in more detail in future work. We also note that small nested droplets that are disappearing during the coarsening process gradually dissolve in a layer-by-layer fashion until they completely vanish (see \href{http://www.princeton.edu/~akosmrlj/videos/}{Video~S3}).

\section{Design of target microstructures}
\label{sec:design}
\begin{figure*}[!t]
\centering
  \includegraphics[width=1.0\textwidth]{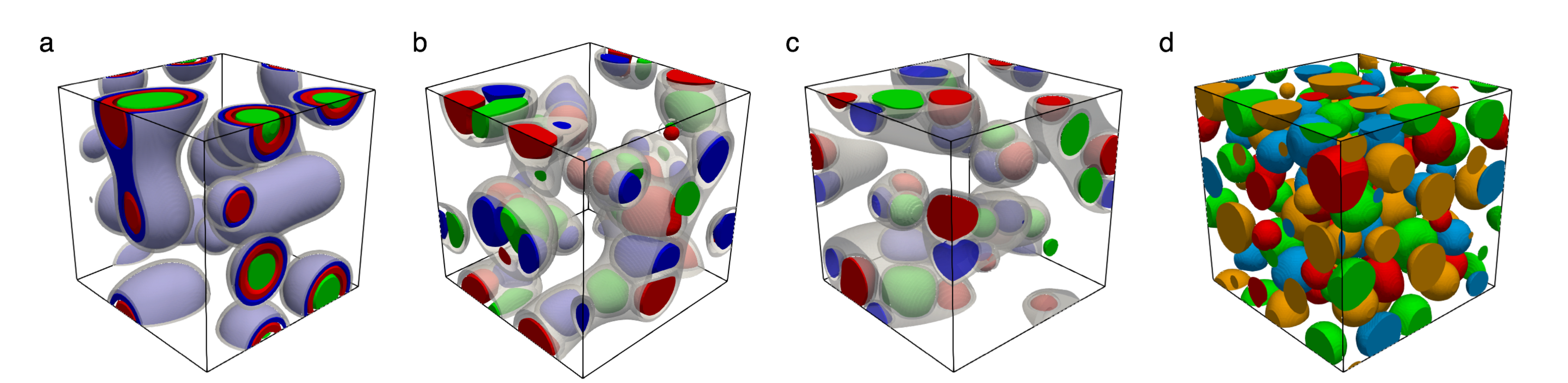}
  \caption{Designed nested morphologies for 5-component mixtures with 5 coexisting phases. (a) ``Russian-doll'' droplets, (b) encapsulated triplets, and (c) encapsulated ``emulsions''. By increasing the volume fraction of the lubricating gray phase in (c), we obtained emulsion with 4 different types of droplets in (d). The majority phase is completely transparent in all panels. The internal structure can be seen in \href{http://www.princeton.edu/~akosmrlj/videos/}{Video~S4, S5, S6, and S7}. Interaction parameters and initial compositions were set to: (a) $\chi_{12} = 2.50,\ \chi_{13} = 5.10,\ \chi_{23} = 2.40,\ \chi_{14} = 6.00,\ \chi_{24} = 5.75, \ \chi_{34} = 2.75, 
 \ \chi_{15} = 7.75, \ \chi_{25} = 7.50, \ \chi_{35}=6.50, \ \chi_{45}=3.00$,  $\lbrace  \bar{\phi}_i \rbrace = \lbrace 0.06, 0.07, 0.08, 0.09, 0.70\rbrace$;
(b) $\chi_{14}=\chi_{15}=\chi_{45}=4.25,\ \chi_{23} = 4.00, \ \chi_{i2} = 3.00,\ \chi_{i3} = 7.00,\ (i = 1,4,5)$,  $\lbrace  \bar{\phi}_i \rbrace = \lbrace 0.067, 0.10, 0.70, 0.066, 0.066\rbrace$;
(c) $\chi_{14}=\chi_{15}=\chi_{45}=6.00,\ \chi_{23} = 4.00, \ \chi_{i2} = 3.00,\ \chi_{i3} = 7.00,\ (i = 1,4,5)$,  $\lbrace  \bar{\phi}_i \rbrace = \lbrace 0.05, 0.15, 0.70, 0.05, 0.05\rbrace$;
(d) $\chi_{14}=\chi_{15}=\chi_{45}=6.00,\ \chi_{23} = 4.00, \ \chi_{i2} = 3.00,\ \chi_{i3} = 7.00,\ (i = 1,4,5)$,  $\lbrace  \bar{\phi}_i \rbrace = \lbrace 0.16, 0.42, 0.15, 0.14, 0.13\rbrace$.
  }
  \label{fig:case5}
\end{figure*}

In this section we discuss how one can rationally design the interaction parameters $\lbrace \chi_{ij} \rbrace$ and average compositions $\lbrace \overline \phi_i \rbrace$ to achieve target microstructures. As was already alluded to in previous sections, the equilibrium microstructure is completely determined from surface tensions between phases and their volume fractions. In general, the relation between the surface tensions and interaction parameters is quite complex. However, it can be drastically simplified in the limit where interaction parameters $\lbrace \chi_{ij} \rbrace$ are large. In this limit the surface tensions are approximately proportional to the interactions parameters [see Eq.~(\ref{eqn:sigma-chi})]. By using this relationship, we discuss how one can rationally design three different microstructures in 5-component mixtures with 5 coexisting phases: `'Russian-doll'' droplets , encapsulated triple ``Janus-like'' droplets, and encapsulated ``micro-emulsions'' (see Fig.~\ref{fig:case5}).

The ``Russian-doll'' droplets with $P=N$ phases $\alpha_1$, $\alpha_2$, \ldots, $\alpha_N$, such that the phases are numbered sequentially with $\alpha_1$ ($\alpha_N$) being the innermost (outermost) phase of the nested structure, require that the surface tensions for an arbitrary triplet $\alpha_i-\alpha_j-\alpha_k$ of phases satisfy the inequality
$\gamma_{\alpha_i \alpha_k} > \gamma_{\alpha_i \alpha_j} + \gamma_{\alpha_j \alpha_k}$, where $i<j<k$. By relating the surface tensions to interaction parameters according to the Eq.~(\ref{eqn:sigma-chi}) and by satisfying these inequalities, we were able to generate the ``Russian-doll'' droplets with 5 phases (see Fig.~\ref{fig:case5}a). Note that the formation of droplets requires that the volume fraction of the outermost phase $\alpha_N$ is sufficiently large to prevent the formation of a nested bicontinuous structure (see Fig.~\ref{fig:case2}b). While we were able to successfully generate ``Russian-dolls'' in this 5-component solution, this might be more challenging in solutions with $N>5$ components within the Flory-Huggins approach. This state of affairs arises due to the fact that the number of inequalities between surface tensions $\binom{N}{3}$ is larger than the number of interaction parameters $\binom{N}{2}$. Thus, it might not be possible to satisfy all the inequalities within the Flory-Huggins model.

Next, we discuss how to design encapsulated triple ``Janus particle''-like droplets, which we refer to as triplets (see Fig.~\ref{fig:case5}b). For simplicity, we assume that the $3$ phases $\alpha$, $\beta$, and $\gamma$, that are forming the triplets, are equivalent, such that their surface tensions $\gamma_{\alpha \beta} = \gamma_{\alpha \gamma} =\gamma_{\beta\gamma}$. The phase $\delta$ that is encapsulating  triplets is shielding them from the surrounding matrix phase $\epsilon$. Therefore, the surface tensions must satisfy the inequalities $\gamma_{\mu\epsilon} > \gamma_{\mu \delta}+ \gamma_{\delta \epsilon}$, where $\mu \in \lbrace \alpha, \beta, \gamma \rbrace$. By satisfying these inequalities and by setting the volume fraction of the matrix phase $\epsilon$ to be sufficiently large, we were indeed able to obtain encapsulated triplets (see Fig.~\ref{fig:case5}b).

Finally, we comment on how to modify interaction parameters to transform the encapsulated triplets to emulsions of $3$ different encapsulated phases. This time, the phase $\delta$ must also shield the phases $\alpha$, $\beta$, and $\gamma$ from each other. The surface tensions thus need to obey the following inequalities: $\gamma_{\mu\nu} > \gamma_{\mu \delta}+ \gamma_{\delta \nu}$, where $\mu,\nu \in \lbrace \alpha, \beta, \gamma, \epsilon \rbrace$. By tuning the volume fractions of individual phases one could obtain either encapsulated emulsions of 3 phases (see Fig.~\ref{fig:case5}c, where $\epsilon$ is the majority phase) or emulsions of 4 phases (see Fig.~\ref{fig:case5}d, where $\delta$ is the majority phase).

To summarize, we demonstrated the first steps towards reverse engineering interaction parameters $\lbrace \chi_{ij} \rbrace$ and average compositions $\lbrace \overline \phi_i \rbrace$ to construct the target microstructure. To specify the morphology of $P$ coexisting phases, there are $\binom{P}{3}$ different inequalities between surface tensions. To ensure that there is enough flexibility, there must be sufficient number of components $N$, such that there are at least as many interaction parameters $\lbrace \chi_{ij} \rbrace$ as there are inequalities between surface tensions $\binom{N}{2} \ge \binom{P}{3}$. The average compositions $\lbrace \overline \phi_i \rbrace$ must be chosen, such that they lie in a region of phase space that correspond to the $P$ coexisting phases with compositions $\lbrace \phi_i^\alpha \rbrace$. By moving the average compositions $\lbrace \overline \phi_i \rbrace$ inside that region one can tune the volume fractions of phases [see Eq.~(\ref{eq:order_parameters2})]. This can be a very complicated task for mixtures with many components and many coexisting phases within the Flory-Huggins model.

\section{Conclusions}
\label{sec:conc}
In this paper we investigated phase diagrams, coarsening and morphologies of 4- and 5-component mixtures. The algorithm developed for constructing phase diagrams based on the convex hull construction of free energy functions is general and can be adapted to an arbitrary physical system with $N\le 5$ components. The PCA and K-Means clustering methods in turn provide convenient means to extract both the number of coexisting phases and their compositions from a given physical realization, also in systems with $N>5$ components that are not directly amenable to phase diagram analysis. 

In agreement with the predictions by Bray~\cite{bray1989exact, bray2002theory}, we found that the coarsening kinetics of multiphase mixtures approaches the $t^{1/3}$ scaling in the long-time limit. However, our data show that phase separation can occur in several stages, and it remains unclear how the coarsening during intermediate stages depends on the number of coexisting phases and their morphology. 

As for the equilibrium packing morphology of coexisting fluid phases, it is completely determined by volume fractions and surface tensions between phases. To this end, we provided guidelines for a rational design of parameters in the Flory-Huggins model that produce target
nested morphologies, such as ``Russian doll'' droplets, encapsulated triplets, and encapsulated emulsions in 5 component systems with 5 coexisting phases.  The design of such structures provides the first steps towards the design of novel self-assembled, autonomic, and hierarchical compartments, that could be used, e.g., for controlled-release systems in medical applications, capable of encapsulating more components than currently achievable with other methods. We note that it might be hard to design arbitrary morphologies in mixtures with more than 5 components within the Flory-Huggins approach, given that the number of inequalities between surface tensions becomes larger than the number of free parameters. This is simply a limitation of the Flory-Huggins model, while other models with more adjustable parameters (or real systems) may provide enough flexibility to achieve the desired structure. 
 
We note that the work reported in this manuscript solely focused on phase separation processes involving spinodal decomposition. At the present time, how nucleation and growth proceeds in multicomponent systems with complex energy landscapes with many local minima and energy barriers, remains an open question. In closing, we hope our work will stimulate further experimental,
numerical, and theoretical investigations of phase behavior and phase transitions in multicomponent systems.

\section*{Acknowledgements}

This research was primarily supported by NSF through the Princeton University's Materials Research Science and Engineering Center DMR-1420541 and through the REU Site EEC-1559973. We would like to acknowledge useful discussions with Mr. Chenyi Fei, Mr. Eliot Feibush, Mr. Yaofeng Zhong and Dr. Lailai Zhu of Princeton University.



\bibliography{library}


\end{document}